\documentclass[journal=mamobx, manuscript=article]{achemso}

\usepackage{eurosym}
\usepackage{enumitem}
\usepackage{tabularx}
\usepackage{graphics}
\usepackage{epsfig}
\usepackage{multirow}
\usepackage{amssymb}
\usepackage{amsmath}
\usepackage{leftidx}
\usepackage{epsfig}
\usepackage{color,soul}
\usepackage{array}
\usepackage{comment}
\usepackage{datetime2}
\usepackage{lscape}
\usepackage{float}
\usepackage{enumitem}

\usepackage[explicit]{titlesec}

\title{Electrostatic Persistence Length Revisited. \\ II. Simulations and Comparison to Experiment}
\date{\today}

\author{Alexey~A.~Gavrilov}
\email{a_gavrilov@ncsu.edu}
\affiliation{Department of Chemical and Biomolecular Engineering, North Carolina State University, Raleigh, North Carolina 27695-7905, United States}

\author{Albert~Johner}
\email{albert.johner@ics-cnrs.unistra.fr}
\affiliation{Institut Charles Sadron, Universit\'{e} de Strasbourg, CNRS UPR22, Strasbourg 67034, France}

\author{Artem~M.~Rumyantsev}
\email{rumyantsev@ncsu.edu}
\affiliation{Department of Chemical and Biomolecular Engineering, North Carolina State University, Raleigh, North Carolina 27695-7905, United States}

\begin{document}

\begin{tocentry}
\includegraphics[height=1.75in]{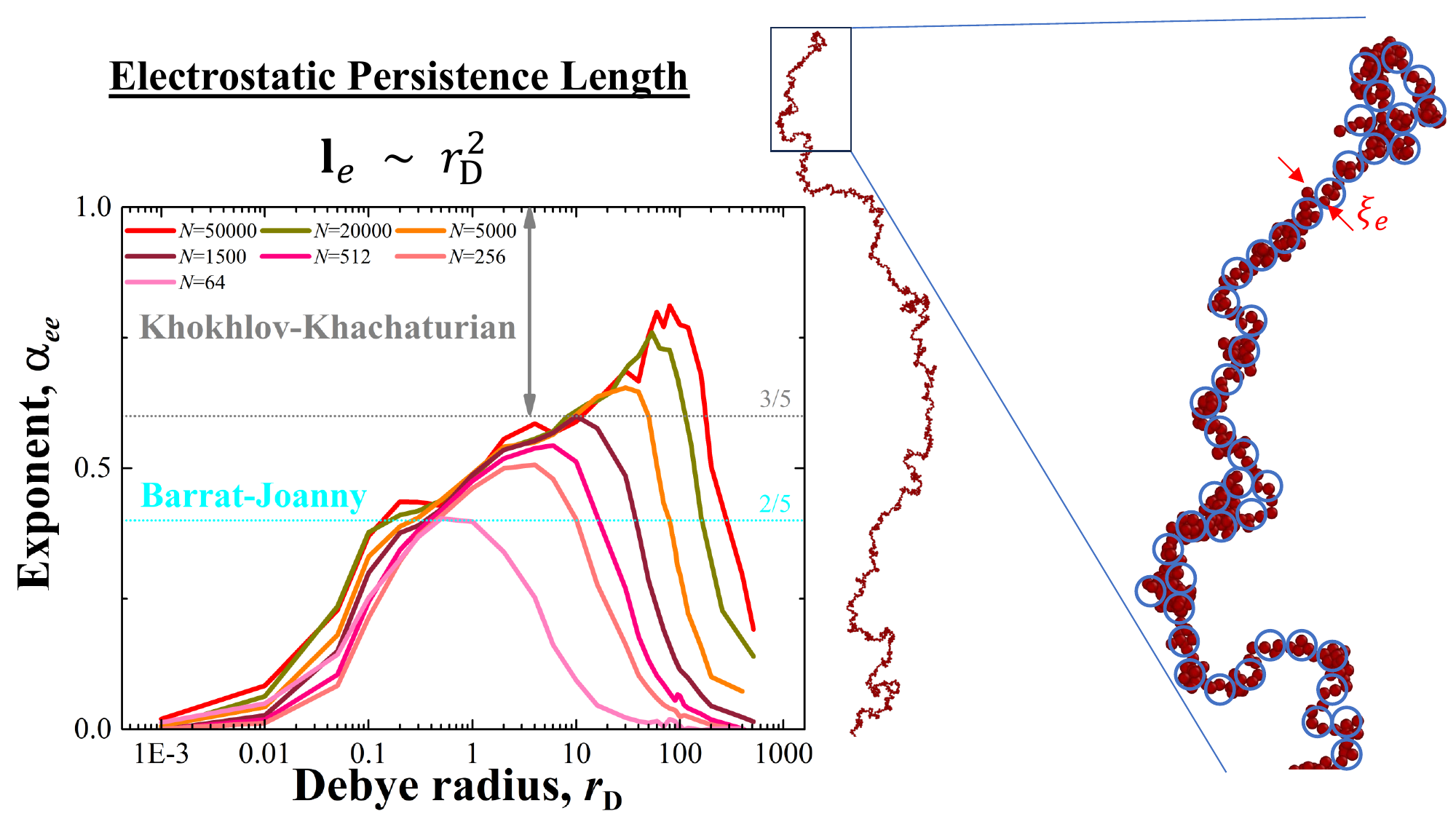}
\end{tocentry}

\begin{abstract}

For decades, debate has surrounded the electrostatic persistence length (EPL) controlling local polyelectrolyte stiffening, centered on two competing power laws: the linear Barrat-Joanny (BJ) prediction, $\textbf{l}_\mathrm{e} \sim r_\mathrm{D}$, and the quadratic Odijk-Skolnick-Fixman/Khokhlov-Khachaturian (OSF/KK) scaling, $\textbf{l}_\mathrm{e} \sim r_\mathrm{D}^{2}$, where $r_\mathrm{D}$ is the Debye screening length. Building on the asymptotic scaling theory developed in the accompanying paper, we validate a complete diagram of limiting regimes using large-scale coarse-grained Monte Carlo simulations of ideal chains with charged monomers interacting through a screened Coulomb potential. By simulating long chains of up to $N \simeq 10^{4} - 10^{5}$ Kuhn segments, we demonstrate that the electrostatic stiffening of both semiflexible and flexible polyelectrolytes obeys the same quadratic OSF/KK law. We track the exponent $\alpha$ which measures how the chain size $R$ grows with the Debye radius, $R \sim r_\mathrm{D}^{\alpha}$. For both cases, in agreement with the OSF/KK theory, $\alpha$ rises past 3/5 and slowly approaches 1 as the chain length increases, whereas within the BJ theory it can never exceed 2/5. We further find that three common size measures, the end-to-end distance $R_\mathrm{ee}$, radius of gyration $R_\mathrm{g}$, and hydrodynamic radius $R_\mathrm{h}$, reach this asymptotic behavior at progressively increasing chain lengths. Consequently, for any real polyelectrolyte, the apparent exponents obey $\alpha_\mathrm{ee} \geq \alpha_\mathrm{g} \geq \alpha_\mathrm{h}$, making $R_\mathrm{g}$ a sharper experimental probe than $R_\mathrm{h}$. Finally, we re-analyze the available experimental data and show that they rule out the linear BJ law and support the quadratic KK scaling, thereby resolving contradictions that stem from mistaking the apparent slopes measured for short chains for the true asymptotic exponent.

\end{abstract}

\maketitle

\newpage


\section{Introduction}
\label{sec:intro}

A complete and coherent understanding of the conformations of single-chain polyelectrolytes (PEs) in salt-added solutions is a fundamental question of the physics of ion-containing polymers. The particularly controversial aspect is the electrostatic stiffening of intrinsically flexible PEs and the associated concept of the electrostatic persistence length (EPL). The formulation of this problem is as follows. The PE chain has a bare Kuhn segment length $l_\mathrm{0}$ and comprises univalent charges equidistantly spaced along the chain at a distance $A = l_\mathrm{0}/f$ from each other. In the presence of implicit univalent salt, these charges repel each other via a screened Coulomb (Yukawa) potential
\begin{equation}
    \frac{U_\mathrm{el}(r)}{k_\mathrm{B} T} = \frac{l_\mathrm{B}}{r} \exp \left( - \frac{r}{r_\mathrm{D}} \right)
\label{eq:Uel-intro}
\end{equation}
where $l_\mathrm{B} = u l_\mathrm{0}$ is the Bjerrum length of the solution and $r_\mathrm{D}$ is the Debye screening radius, which is inversely proportional to the square root of the salt ions concentration, $r_\mathrm{D} \sim c_\mathrm{s}^{-1/2}$. The PE chain is ideal, meaning that all other non-electrostatic interactions between the monomers are absent. 

In the first manuscript of this series,~\cite{EPL-part1} a scaling theory of single-chain PEs was developed, and the analogy between intrinsically stiff/semiflexible chains with $l_\mathrm{B} l_\mathrm{0} /A^{2} = uf^{2} \gg 1$ and intrinsically flexible counterparts with $l_\mathrm{B} l_\mathrm{0}/A^{2} = uf^{2} \ll 1$ was explored and revealed in depth. This suggested the validity of the squared (so-called Khokhlov-Khachaturian,~\cite{KK-1982} KK) scaling of the electrostatic persistence length with the Debye radius, $\textbf{l}_\mathrm{KK} \sim r_\mathrm{D}^{2}$, which is analogous to the squared (Odijk-Skolnick-Fixman,\cite{odijk-1977, SF-1977} OSF) scaling for semiflexible chains. In the present, second part of our comprehensive study, we perform extensive Monte Carlo simulations of a minimal model of a single-chain PE in a salt-added solution in order to comprehensively validate these theoretical results. We note that these simulations are largely aimed at testing the predictions of the standard theoretical model for ideal PEs in salt-added solutions, where counterions and salt are treated implicitly through the effective interaction potential of eq.~\ref{eq:Uel-intro}. Consequently, explicit ion effects (e.g., salt and counterion type) and backbone solvent quality lie beyond the scope of this study.

The earlier simulation studies of these models,~\cite{barrat-1993, kremer-1997, ullner-2002, everaers-2002, shklovskii-2002, dobrynin-2009, DC-2009, prochazka-2012, buehler-2012, stevens-2018} did provide some advance but did not completely clarify all aspects of the problem. Part of the reason for that, as we demonstrate below, was the limitations in computing resources, which restricted the maximal chain length, i.e., the number of bare Kuhn segments $N = L/l_\mathrm{0}$, to (in the context of the EPL problem) \textit{only} several thousand.

In particular, Nguyen and Shklovskii~\cite{shklovskii-2002} studied chains with the bare persistence length equal to the inter-charge spacing equal to the solution Bjerrum length, $l_\mathrm{0} = l_\mathrm{B} = A = 1$, i.e., $uf^{2} = 1$, and $N$ up to 4096. While it was shown that the dependence of the chain end-to-end distance $R_\mathrm{ee}$ on the Debye radius $r_\mathrm{D}$ is more consistent with the OSF/KK scaling of the EPL, these parameter values (as we demonstrate below) correspond to the crossover between stiff and flexible chains and cannot be considered an unambiguous proof of the KK scaling for flexible PEs.

In the other study, Everaers, Milchev, and Yamakov~\cite{everaers-2002} considered the chains of the same $N = 4096$ length but a much lower Bjerrum length and hence $uf^{2} \ll 1$, corresponding to flexible (synonymously, weakly charged) PEs. Representing their data in theoretically-informed coordinates (parameters) revealed that the KK theory provides an imperfect but better data collapse compared to the alternative Barrat-Joanny (BJ) theory~\cite{BJ-1993} for flexible PEs, which suggests a linear scaling of the EPL with the Debye radius, $\textbf{l}_\mathrm{BJ} \sim r_\mathrm{D}$. It was therefore concluded that the linear BJ scaling and other sublinear dependencies are ruled out, but at the same time no direct positive confirmation of the KK scaling slopes was provided.

Thus, in the present work, we aim to reach the following key goals. First, we will demonstrate the validity of the scaling regimes, i.e., the scaling slopes $\alpha$ in the $R_\mathrm{ee} \sim r_\mathrm{D}^{\alpha}$, which are unique to the KK theory. Second, we corroborate the microscopic (``blob'') picture describing conformations of flexible PEs suggested by KK theory and elaborated in the first part of the work.~\cite{EPL-part1} Finally, we elucidate the role of finite-length effects by providing the characteristic chain length $N$, above which the asymptotic KK scaling exponents start to be well pronounced and clearly distinguishable from those predicted within the BJ approach.

The paper is organized as follows. In Section~\ref{sec:model}, we describe the simulation setup and procedure. Section~\ref{sec:theory-recap} provides a brief summary of the theoretical results presented in the first part of this series~\cite{EPL-part1} and specifies the regions of the diagram that will be tested in simulations. Section~\ref{sec:results} presents the simulation results. We start by confirming the validity of the KK quadratic scaling for flexible PEs and then extend the analysis to semiflexible chains, confirming the OSF result. In Section~\ref{sec:experiment}, we demonstrate that our theoretical and simulation results are also supported by the available experimental data for the dependence of the chain size on the salt concentration: they favor the squared KK scaling and completely rule out the linear BJ alternative for the EPL. Conclusions are summarized in Section~\ref{sec:conclusions}.

\section{Model and Method}
\label{sec:model}

We model a single linear polyelectrolyte (PE) chain with a bare Kuhn segment length $l_\mathrm{0}$ in continuous three-dimensional space. The chain carries charged monomers spaced equidistantly at the curvilinear distance $A$ from each other. The chain contour length is equal to $L = l_\mathrm{0} N$, where $N$ is the number of Kuhn segments in the chain. Consecutive monomers are connected by bonds of fixed length $b$. The number of monomers in the chain is therefore equal to $L/b$, and the fraction of charged monomers is $b/A$. Because the bonds are inextensible, the energy contains no bond-stretching contribution, and the chain conformations are determined entirely by a bending energy and a screened electrostatic interaction between monomers. The bond length $b$ is taken as the unit of length (i.e., $b = 1.0$), and all lengths ($L, l_\mathrm{0}, l_\mathrm{B}, A$) will be expressed in units of $b$ for brevity in what follows. The energies are expressed in units of the thermal energy $k_\mathrm{B} T$ ($\beta = 1/k_\mathrm{B} T$).

Charged monomers interact through a Yukawa (screened Coulomb) potential,
\begin{equation}
    \beta U_\mathrm{el}(r) = \frac{l_\mathrm{B}}{r} \exp \left( - \frac{r}{r_\mathrm{D}} \right)
\label{eq:Uel}
\end{equation}
where $r$ is the inter-monomer separation, $l_\mathrm{B}$ is the Bjerrum length, and $r_\mathrm{D}$ is the Debye screening length. Chain stiffness is introduced through a bending potential acting on each pair of consecutive bonds,
\begin{equation}
    \beta U_\mathrm{b}(\theta_i) = K (1 - \cos \theta_i)
\label{eq:Ub}
\end{equation}
where $\theta_i$ is the angle between successive bonds and $K$ is the bending rigidity, which sets the backbone bare Kuhn segment length $l_{0}$. The bare Kuhn segment length is calculated from $K$ using the following expression:~\cite{carrillo-2013}
\begin{equation}
    l_\mathrm{0} = \frac{1 + \coth (K) - K^{-1}}{1 - \coth (K) + K^{-1}}
\label{eq:l0-from-K}
\end{equation}
The total potential energy is the sum of eq.~\ref{eq:Uel} over all non-bonded pairs and of eq.~\ref{eq:Ub} over all interior monomers.

Equilibrium conformations were sampled by Metropolis Monte Carlo using the pivot algorithm,~\cite{everaers-2002, shklovskii-2002, madras-1988} which provides efficient relaxation of the global conformation of a single chain. In each move a monomer is chosen at random as the pivot point, and all monomers on one side of it are rotated rigidly about the pivot point by a random rotation, with the rotation axis drawn uniformly on the unit sphere and the rotation angle uniformly in $[0, 2\pi]$. Being a rigid-body rotation, the move preserves all bond lengths and the internal structure of the displaced segment, so that only the interaction between the two segments and the bond angle at the pivot point are altered. Each trial conformation was accepted with the Metropolis probability $\min [1, \exp(-\beta \Delta U)]$. A Monte Carlo step consisted of $L/b$ attempted pivot moves; the chain was first equilibrated from an extended initial conformation and then sampled over a long production run consisting of at least 1000 steps.

Three quantities were used in our simulations to characterize the chain size: average squared end-to-end distance $\left\langle R_\mathrm{ee}^{2} \right\rangle$, average squared radius of gyration $\left\langle R_\mathrm{g}^{2} \right\rangle$, and average hydrodynamic radius $\left\langle R_\mathrm{h} \right\rangle$. They were calculated using the following expressions:~\cite{GK-book}
\begin{equation}
    \left\langle R_\mathrm{ee}^{2} \right\rangle = \left\langle \left( \mathbf{r}_N - \mathbf{r}_1 \right)^{2} \right\rangle
\label{eq:Ree2}
\end{equation}
\begin{equation}
    \left\langle R_\mathrm{g}^{2} \right\rangle = \frac{1}{2N^{2}} \left\langle \sum_{i,j} \left( \mathbf{r}_j - \mathbf{r}_i \right)^{2} \right\rangle
\label{eq:Rg2}
\end{equation}
\begin{equation}
    \left\langle R_\mathrm{h} \right\rangle^{-1} = \frac{1}{N^{2}} \left\langle \sum_{i,j \, (i \neq j)} \left| \mathbf{r}_j - \mathbf{r}_i \right|^{-1} \right\rangle
\label{eq:Rh}
\end{equation}
where the angular brackets represent an ensemble average. To enable direct comparison of the simulation results with the scaling exponents predicted theoretically~\cite{EPL-part1} and measured in experiments, we use the root-mean-square end-to-end distance and radius of gyration, for brevity denoted $R_\mathrm{ee} \equiv \sqrt{\left\langle R_\mathrm{ee}^{2} \right\rangle}$ and $R_\mathrm{g} \equiv \sqrt{\left\langle R_\mathrm{g}^{2} \right\rangle}$; similarly, $R_\mathrm{h}$ is the average hydrodynamic radius $\left\langle R_\mathrm{h} \right\rangle$.

\section{Theoretical Diagram and Regimes Recap}
\label{sec:theory-recap}

\begin{figure}
\centering
\includegraphics[width=6.5in]{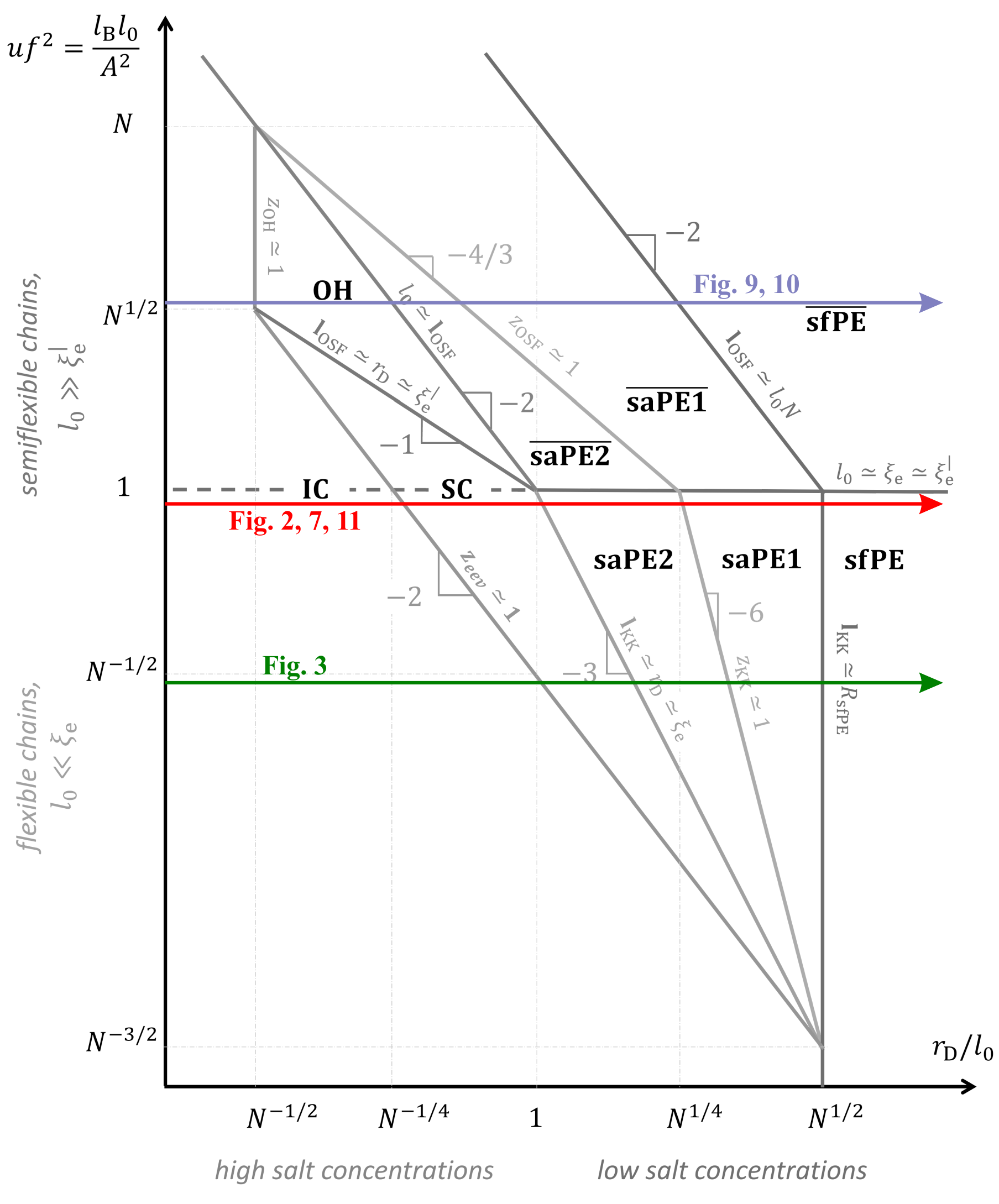}
\caption{Theoretical scaling diagram constructed in the accompanying theoretical work~\cite{EPL-part1}, with simulation tests marked as colored arrows. Each simulation test refers to the corresponding figure in the present text.}
\label{fig:1}
\end{figure}

In this section we will briefly discuss the theoretical findings presented in the first part of the series~\cite{EPL-part1} to establish the parameter mapping as well as important scaling laws to be tested in simulations. We begin with the main scaling diagram presented in Figure~\ref{fig:1} in the coordinates $uf^{2}$ versus $r_\mathrm{D}/l_\mathrm{0}$. These universal parameters provide the only way to represent the solution of a four- or even five-parameter problem (with $l_\mathrm{0}, l_\mathrm{B}, A,$ $r_\mathrm{D}$ and $L$ lengths) in terms of a conventional two-dimensional diagram of scaling regimes. They are the reduced Coulomb strength, $uf^{2} = l_\mathrm{B} l_\mathrm{0}/A^{2}$, equal to the electrostatic self-energy of a bare Kuhn segment, and the reduced Debye radius, $r_\mathrm{D}/l_\mathrm{0}$, which quantifies the degree of screening of the Coulomb repulsions. It should be emphasized once again that the parameter $N$ in the theory, and hence in the simulations, is not the number of chemical monomers but rather the number of bare Kuhn segments in the chain, $N = L/l_\mathrm{0}$. In what follows, for $l_\mathrm{0} \neq 1$ in simulations, we report both the number of beads (chemical monomers) equal to the contour length, $L/b = L$, and the number of segments, $N = L/l_\mathrm{0}$. Table~\ref{table:1} provides the scaling laws for the chain sizes for each of the conformational regimes identified in Figure~\ref{fig:1}. The laws expressed in terms of the salt concentration $c_\mathrm{s}$ can be found in the first paper of the series.~\cite{EPL-part1}

\begin{center}
\begin{table*} 
\caption{Scaling laws for polyelectrolyte chain size $R$ in different conformational regimes.}
\label{table:1}
\centering
\renewcommand{\arraystretch}{2.0}
\begin{tabular}{|c|c|c|}
\hline
Regime
&  Conformation type
&  Chain size $R$
\\ \hline
IC
&  ideal coil
&  $ l_\mathrm{0}^{1/2} L^{1/2} $
\\ \hline
SC
&  swollen coil
&  $ l_\mathrm{0}^{1/5} l_\mathrm{B}^{1/5} A^{-2/5} r_\mathrm{D}^{2/5} L^{3/5} $
\\ \hline
saPE2
&  globally swollen coil
&  $ l_\mathrm{0}^{1/15} l_\mathrm{B}^{4/15} A^{-8/15} r_\mathrm{D}^{3/5} L^{3/5} $
\\ \hline
saPE1
&  globally ideal coil
&  $ l_\mathrm{0}^{-1/6} l_\mathrm{B}^{1/3} A^{-2/3} r_\mathrm{D} L^{1/2} $
\\ \hline
sfPE
&  rodlike stretch
&  $ l_\mathrm{0}^{1/3} l_\mathrm{B}^{1/3} A^{-2/3} L $
\\ \hline
OH
&  globally swollen coil
&  $ l_\mathrm{0}^{1/5} r_\mathrm{D}^{1/5} L^{3/5} $
\\ \hline
$\overline{\text{saPE2}}$
&  globally swollen coil
&  $ l_\mathrm{B}^{1/5} A^{-2/5} r_\mathrm{D}^{3/5} L^{3/5} $
\\ \hline
$\overline{\text{saPE1}}$
&  globally ideal coil
&  $ l_\mathrm{B}^{1/2} A^{-1} r_\mathrm{D} L^{1/2} $
\\ \hline
$\overline{\text{sfPE}}$
&  rodlike stretch
&  $ L $
\\ \hline
\end{tabular}
\end{table*}
\end{center}

Since in simulations the chain length $N$ is a natural parameter, we will also consider how the positions of the crossovers between neighboring regimes evolve with $N$. The most convenient way to test theoretical predictions is to explore the dependence of the chain size at the crossover, $R_\mathrm{ee}^\mathrm{X/Y}$, on the very position of the crossover along the Debye radius coordinate, $r_\mathrm{D}^\mathrm{X/Y}$, across different chain lengths $N$. Theory developed in the first part of the series enables deriving the respective dependences, $R_\mathrm{ee}^\mathrm{X/Y}(r_\mathrm{D}^\mathrm{X/Y})$, by explicitly obtaining both quantities and then eliminating the parameter $N$ from the resulting parametric dependence. Having such relations summarized in Table~\ref{table:2} for the adjacent scaling regimes of flexible chains (cf.\ Figure~\ref{fig:1}), one can trace how the X/Y crossover on a single $R_\mathrm{ee}(r_\mathrm{D})$ plot shifts upon varying $N$.

\begin{center}
\begin{table*} 
\caption{Dependence of the crossover chain size $R_\mathrm{ee}^\mathrm{X/Y}$ on the crossover coordinate $r_\mathrm{D}^\mathrm{X/Y}$.}
\label{table:2}
\centering
\renewcommand{\arraystretch}{2.0}
\begin{tabular}{|c|c|}
\hline
Crossover
&  Scaling expression
\\ \hline
IC/SC
&  $ R_\mathrm{ee}^\mathrm{IC/SC} \sim \left( r_\mathrm{D}^\mathrm{IC/SC} \right)^{-2} $
\\ \hline
SC/saPE2
&  $ R_\mathrm{ee}^\mathrm{SC/saPE2} \sim \left( r_\mathrm{D}^\mathrm{SC/saPE2} \right)^{0} $
\\ \hline
saPE2/saPE1
&  $ R_\mathrm{ee}^\mathrm{saPE2/saPE1} \sim \left( r_\mathrm{D}^\mathrm{saPE2/saPE1} \right)^{3} $
\\ \hline
saPE1/sfPE
&  $ R_\mathrm{ee}^\mathrm{saPE1/sfPE} \sim \left( r_\mathrm{D}^\mathrm{saPE1/sfPE} \right)^{3} $
\\ \hline
saPE2/sfPE
&  $ R_\mathrm{ee}^\mathrm{saPE2/sfPE} \sim \left( r_\mathrm{D}^\mathrm{saPE2/sfPE} \right)^{3/2} $
\\ \hline
\end{tabular}
\end{table*}
\end{center}

We note that the last crossover, saPE2/sfPE, is absent for sufficiently long chains, when the scaling regime saPE1 (intermediate between these two) fully develops, as shown in Figure~\ref{fig:1}. However, for short chains, the saPE1 regime window is so narrow that it is effectively absent: the numerical prefactors, which differ from unity and cannot be specified by scaling arguments, shift its boundaries. For this reason, the saPE2/sfPE crossover is observed for short chains instead of the consecutive saPE2/saPE1 and saPE1/sfPE ones.

Beyond the global chain size, we will also investigate the internal chain statistics. Namely, we will test the scaling picture of flexible PEs ($uf^{2} \ll 1$) exhibiting KK stiffening. This implies that, locally, the chain can be viewed as a rodlike array of Gaussian ($\Theta$-solvent) electrostatic blobs, each of the size $\xi_\mathrm{e}$ and containing $g_\mathrm{e}$ Kuhn segments given by
\begin{equation}
    \xi_\mathrm{e} \simeq l_\mathrm{0} (uf^{2})^{-1/3}; \quad g_\mathrm{e} \simeq \left( \frac{\xi_\mathrm{e}}{l_\mathrm{0}} \right)^{2} \simeq (uf^{2})^{-2/3}
\label{eq:xie-ge}
\end{equation}
We will also pay particular attention to the internal mean-square distances of flexible PEs in the saPE2 regime, which exhibits the most complex behavior owing to the larger number of relevant length scales: the electrostatic blob size $\xi_\mathrm{e}$, the KK EPL $\textbf{l}_\mathrm{e}$, the thermal blob size of the KK quasi-monomers $\xi_\mathrm{T}^\mathrm{KK}$, and the total chain size $R_\mathrm{saPE2}$. Detailed physical derivation of this physical picture can be found in the theoretical part of the series,~\cite{EPL-part1} and Figure S1 of the Supporting Information shows the predicted behavior of the internal MSD. In brief, at increasing separation between the monomers, internal chain statistics in the saPE2 regime exhibits four distinct consecutive behaviors: (i) Gaussian behavior with $R_\mathrm{ij}^{2} \sim |i-j|$ within the electrostatic blob; (ii) rodlike statistics $R_\mathrm{ij}^{2} \sim |i-j|^{2}$ at the length between the electrostatic blob and the KK EPL; (iii) again Gaussian statistics $R_\mathrm{int}^{2} \sim |i-j|$ due to the ideal random walk of rodlike KK quasi-monomers within the thermal blob $\xi_\mathrm{T}^\mathrm{KK}$; (iv) finally, swollen coil statistics $R_\mathrm{int}^{2} \sim |i-j|^{6/5}$ at the largest lengths due to the self-avoiding random walk of the thermal blobs.

\subsection{Parameter Limitations}
\label{sec:param-limits}

Here we discuss the parameter limitations important for the simulation setup. It seems appealing to decrease the number of charges on the chain (i.e., increase their spacing $A$) and simultaneously increase the Bjerrum length $l_\mathrm{B}$ to keep the universal parameter of electrostatic strength $uf^{2} = l_\mathrm{B} l_\mathrm{0}/A^{2}$ fixed. However, the theoretical derivations assumed (for flexible regimes with electrostatic stiffening saPE2, saPE1, sfPE, and their semiflexible counterparts $\overline{\text{saPE2}}$, $\overline{\text{saPE1}}$, $\overline{\text{sfPE}}$) that the discrete charges are replaced by a continuously smeared charge density. This assumption is valid when $A \ll r_\mathrm{D}$ for semiflexible chains (i.e., for $uf^{2} \gg 1$), and when $(A l_\mathrm{0} l_\mathrm{B})^{1/3} \ll r_\mathrm{D}$ for flexible chains (i.e., for $uf^{2} \ll 1$).~\cite{EPL-part1, GK-book} Since the condition is rather strict for semiflexible PEs, we adhere to $A = 1$, i.e., each chemical monomer (bead) carries a charge, when simulating systems at $uf^{2} \geq 1$. For flexible chains, however, the restriction is much weaker, and, when the Bjerrum length $l_\mathrm{B}$ is low, the fraction of charged monomers $f = b/A = 1/A$ can be safely reduced to improve the computational efficiency. To determine the $r_\mathrm{D}$-range where using higher charge spacings $A$ is safe and does not affect the system behavior, we performed preliminary simulations at $uf^{2} = 1/32$ (being the largest studied $uf^{2}$ value in the flexible chain region) for three different values of $A = 1$, 4, and 16. The resulting $R_\mathrm{ee}(r_\mathrm{D})$ dependencies presented in Figure S2 of the Supporting Information are identical for $A = 1$ and 4 for $r_\mathrm{D} \geq 16$. Therefore, when studying flexible PE with $uf^{2} \ll 1$, we chose $A = 4$ but restrict ourselves to $r_\mathrm{D} \geq 16$.

It should be noted that the requirement of $A \ll r_\mathrm{D}$ needs to be fulfilled in the OH regime of semiflexible chains as well, ensuring that the Debye blobs of adjacent charges overlap and form around the chain a continuous excluded-volume cylindrical shell.~\cite{EPL-part1}

\section{Results and Discussions}
\label{sec:results}

\subsection{Polyelectrolytes at the Flexible/Semiflexible Crossover}
\label{sec:crossover-flex-semiflex}

\begin{figure} 
\centering
\includegraphics[width=3.25in]{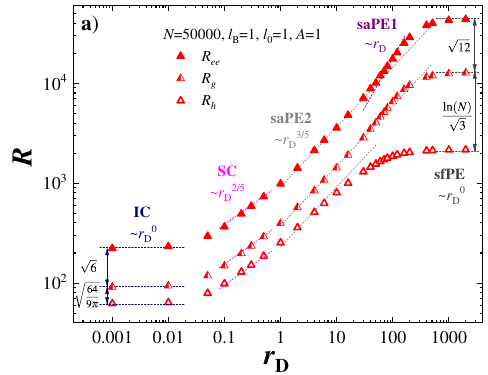} \\ 
\includegraphics[width=3.25in]{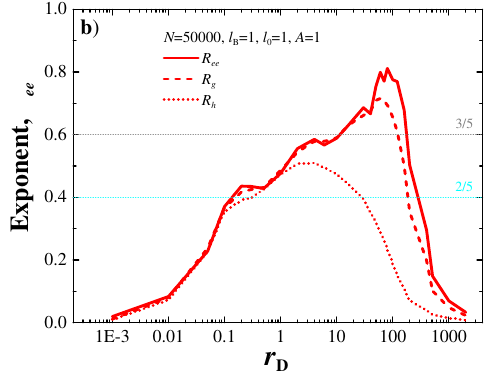}
\caption{(a) Dependencies of the polyelectrolyte end-to-end distance $R_\mathrm{ee}$, gyration radius $R_\mathrm{g}$, and hydrodynamic radius $R_\mathrm{h}$ on the Debye screening radius $r_\mathrm{D}$ for $N = 50000$, $l_\mathrm{B} = 1$, $A = 1$, $l_\mathrm{0} = 1$, and $L = N l_\mathrm{0} = 50000$. (b) Logarithmic derivative of this dependence, $\alpha_\mathrm{ee} = \partial(\ln R)/\partial(\ln r_\mathrm{D})$, corresponding to the apparent slope in the $R \sim r_\mathrm{D}^{\alpha}$ law.}
\label{fig:2}
\end{figure}

We start with the most straightforward case of $l_\mathrm{B} = 1$, $A = 1$, and $l_\mathrm{0} = 1$ corresponding to $uf^{2} = 1$, and explore the dependence of the PE size on the Debye radius $r_\mathrm{D}$. It will be later shown that this set of parameters corresponds to PEs close to the flexible/semiflexible crossover region on the scaling diagram (Figure~\ref{fig:1}), hence changing the $r_\mathrm{D}$ value corresponds to moving along the red arrow in it. The obtained dependencies of the different metrics of the chain size ($R_\mathrm{ee}$, $R_\mathrm{g}$, and $R_\mathrm{h}$) on the Debye radius $r_\mathrm{D}$ are presented in Figure~\ref{fig:2}a.

Let us first consider the end-to-end distance $R_\mathrm{ee}$ and gyration radius $R_\mathrm{g}$ behavior. The main observation is that all 5 expected regimes (IC, SC, saPE2, saPE1, and sfPE) are recovered in the theoretically predicted order, cf.\ Figure~\ref{fig:1}. At very low $r_\mathrm{D}$-values (i.e., very high salt concentrations), the chain behaves as a neutral chain, and its size does not depend on $r_\mathrm{D}$ -- this is the ideal-coil regime IC. As $r_\mathrm{D}$ increases, the chain enters the swollen coil regime SC, where its size increases with the Debye radius as $R \sim r_\mathrm{D}^{2/5}$ (see Table~\ref{table:1}). In the next regime, saPE2, where the chain behavior starts to be influenced by the electrostatic stiffening, the size increases as $R \sim r_\mathrm{D}^{3/5}$. Upon further increasing $r_\mathrm{D}$, the chain size scaling with $r_\mathrm{D}$ becomes even steeper as it tends toward the slope of unity, $R \sim r_\mathrm{D}^{1}$, predicted for the saPE1 regime. Finally, when the $r_\mathrm{D}$-values become very high, there is essentially no salt screening (i.e., the salt concentration is very low), and in the sfPE regime the chain behaves as a salt-free PE with its size independent of $r_\mathrm{D}$.

In order to investigate the observed regimes in more detail, the $r_\mathrm{D}$-dependence of the exponent $\alpha$ in the dependence $R \sim r_\mathrm{D}^{\alpha}$ is presented in Figure~\ref{fig:2}b. To extract the apparent slope value $\alpha$, the logarithmic derivative of the $R(r_\mathrm{D})$ dependence was calculated as $\alpha = \partial \ln(R)/\partial \ln(r_\mathrm{D})$. At very low and very high $r_\mathrm{D}$, the value of $\alpha$ tends to zero, in agreement with the scaling laws $R \sim r_\mathrm{D}^{0}$ for the IC and sfPE regimes, see Table~\ref{table:1}. At intermediate values of the Debye radius $r_\mathrm{D}$, the exponent $\alpha$ exhibits two plateaus at $\alpha \approx 2/5$ and $\alpha \approx 3/5$, corresponding to the theoretical regimes SC and saPE2, respectively. As $r_\mathrm{D}$ increases further, the system crosses over to regime saPE1 and the value of $\alpha$ keeps growing until it reaches its maximum of approximately 0.82 for $R_\mathrm{ee}$ and 0.73 for $R_\mathrm{g}$. It is interesting that these values are still lower than $\alpha = 1$ predicted in theory, cf.\ Table~\ref{table:1} for the saPE1 regime. It will be shown below in Section~\ref{sec:chain-length} that this discrepancy is related to the fact that even for $N = 50000$ the saPE1 regime is not yet fully developed, and even longer chains are necessary for $\alpha$ to fully reach the theoretically predicted asymptotic value of exponent $\alpha = 1$.

Let us now address the differences in the observed apparent scaling exponents $\alpha_\mathrm{ee}$ and $\alpha_\mathrm{g}$ for $R_\mathrm{ee}$ and $R_\mathrm{g}$, as well as the seemingly drastically different behavior of the $\alpha_\mathrm{h}$ slope for the hydrodynamic radius $R_\mathrm{h}$. Figure~\ref{fig:2}b shows that, up to $r_\mathrm{D} \approx 1$, the observed slope is the same for all three chain size characteristics, $\alpha_\mathrm{ee} \approx \alpha_\mathrm{g} \approx \alpha_\mathrm{h}$. At higher $r_\mathrm{D}$-values, however, the exponent $\alpha_\mathrm{h}$ for $R_\mathrm{h}$ becomes systematically lower compared to those for both $R_\mathrm{ee}$ and $R_\mathrm{g}$. At $r_\mathrm{D} > 20$, the same happens to the exponent $\alpha_\mathrm{g}$ extracted from $R_\mathrm{g}$ as it becomes smaller than $\alpha_\mathrm{ee}$, and the hierarchy $\alpha_\mathrm{ee} \geq \alpha_\mathrm{g} \geq \alpha_\mathrm{h}$ becomes noticeable.

The reason for such behavior lies in the fact that, upon passing through all the consecutive regimes, the chain undergoes a form-factor change. Initially, at the lowest $r_\mathrm{D}$, the PE is an ideal coil (regime IC), and finally, at the largest $r_\mathrm{D}$, it is rod-like (regime sfPE). This can be seen in Figure~\ref{fig:2}a, where
\begin{equation}
    \left( \frac{R_\mathrm{ee}}{R_\mathrm{g}} \right)_\mathrm{IC} = \sqrt{6}; \qquad \left( \frac{R_\mathrm{g}}{R_\mathrm{h}} \right)_\mathrm{IC} = \sqrt{\frac{64}{9\pi}}
\label{eq:ratio-IC}
\end{equation}
for chains in the IC regime and
\begin{equation}
    \left( \frac{R_\mathrm{ee}}{R_\mathrm{g}} \right)_\mathrm{rod} = \sqrt{12}; \qquad \left( \frac{R_\mathrm{g}}{R_\mathrm{h}} \right)_\mathrm{rod} = \frac{\ln(N)}{\sqrt{3}}
\label{eq:ratio-rod}
\end{equation}
for globally rod-like conformations in the sfPE regime. The $\sqrt{12}/\sqrt{6} = \sqrt{2}$ factor results in the gyration radius $R_\mathrm{g}$ growing slower than the end-to-end distance $R_\mathrm{ee}$ upon increasing $r_\mathrm{D}$. In turn, $R_\mathrm{h}$ demonstrates an even slower growth compared to $R_\mathrm{g}$, with a factor of $(\ln(N)/\sqrt{3})/\sqrt{64/9\pi} = \sqrt{3\pi}\ln(N)/8$ accumulating over the entire range of $r_\mathrm{D}$, which increases with the chain length $N$. Therefore, the $R_\mathrm{g}$ and especially $R_\mathrm{h}$ dependences on $r_\mathrm{D}$ can be considered as underestimating the scaling exponents. Alternatively, one may argue that longer PE lengths are required for $R_\mathrm{g}$ and especially $R_\mathrm{h}$ to reach the asymptotic theoretical slopes of $\alpha$ compared to $R_\mathrm{ee}$. Therefore, in what follows, we focus on $R_\mathrm{ee}$ as it is better suited for determining the scaling regimes, since it demonstrates a sharper increase upon increasing $r_\mathrm{D}$.

It is also worth pointing out here that the exponent for $R_\mathrm{h}$ does not actually reach the level of even $\alpha \approx 3/5$ (see Figure~\ref{fig:2}b) despite $N = 50000$. Since $R_\mathrm{h}$ is the quantity frequently investigated in experimental studies, this fact will be important later when we discuss the literature data in Section~\ref{sec:experiment}.

We finally note that the scaling slopes versus $r_\mathrm{D}$ are the same in the intermediate regimes for flexible and semiflexible PEs: 2/5 for regime SC common in both cases, 3/5 for flexible regime saPE2 and semiflexible $\overline{\text{saPE2}}$, and 1 for saPE1 and $\overline{\text{saPE1}}$ regimes (see Figure~\ref{fig:1} and Table~\ref{table:1}). Therefore, the above results obtained at the flexible/semiflexible PE crossover, $uf^{2} = 1$, are consistent with both OSF theory for semiflexible chains and KK theory for flexible chains. However, one may consider them insufficient to completely prove the quadratic KK scaling, $\textbf{l}_\mathrm{e} \sim r_\mathrm{D}^{2}$ (and therefore rule out the linear BJ alternative of $\textbf{l}_\mathrm{e} \sim r_\mathrm{D}^{1}$) pertaining to substantially flexible chains, $uf^{2} \ll 1$. To this end, in the next subsection, we focus on that case.

\subsection{Flexible Polyelectrolytes}
\label{sec:flexible-PE}

Next, we move down on the diagram of Figure~\ref{fig:1} (see the green arrow there) to study the case of weaker electrostatic interactions, i.e., the region of flexible PEs. Since regimes IC and SC observed at low $r_\mathrm{D}$ are not of significant interest, as the chain is not electrostatically stiffened in them, we restrict ourselves to $r_\mathrm{D} \geq 16$. This allows us to increase the charged monomer spacing to $A = 4$ (i.e., $f = 1/4$, see Section~\ref{sec:param-limits}) for better computational efficiency, and, subsequently, to study chains of the length $N = 100000 = 10^{5}$. The increased chain length is necessary because, according to the theoretical diagram as well as preliminary runs for $N = 20000$, we expect all regimes (except for the trivial IC and sfPE) to become narrower as the chains become more flexible, that is, as $uf^{2}$ goes down (see Figure~\ref{fig:1}).

\begin{figure}
\centering
\includegraphics[width=3.25in]{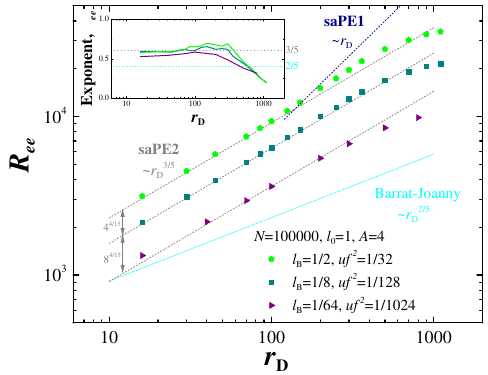}
\caption{The dependence of the end-to-end distance $R_\mathrm{ee}$ of flexible (i.e., weakly charged) polyelectrolyte on the Debye radius $r_\mathrm{D}$ for three values of the Coulomb strength parameter $uf^{2} = 1/32$, 1/128, and 1/1024. PEs comprise $N = 100000$ bare statistical segments of length $l_\mathrm{0} = 1$, so the chain contour length is $L = N l_\mathrm{0} = 100000$. The inset shows the corresponding logarithmic derivative dependencies, $\alpha_\mathrm{ee} = \partial(\ln R)/\partial(\ln r_\mathrm{D})$.}
\label{fig:3}
\end{figure}

Three values of the Coulomb strength parameter were investigated: $uf^{2} = 1/32$ (by setting $l_\mathrm{B} = 1/2$, $A = 4$, $l_\mathrm{0} = 1$), $uf^{2} = 1/128$ (using $l_\mathrm{B} = 1/8$, $A = 4$, $l_\mathrm{0} = 1$), and $uf^{2} = 1/1024$ (achieved by using $l_\mathrm{B} = 1/64$, $A = 4$, and $l_\mathrm{0} = 1$). Figure~\ref{fig:3} shows the resulting dependencies of $R_\mathrm{ee}$ on $r_\mathrm{D}$.

As one can see, the saPE2 scaling $R \sim r_\mathrm{D}^{3/5}$ is nicely reproduced in a wide range of $r_\mathrm{D}$ values for $uf^{2} = 1/32$ and $1/128$, which can clearly be seen from the logarithmic derivative $\alpha$ dependence provided in the inset of Figure~\ref{fig:3}. For $uf^{2} = 1/1024$, the exponent $\alpha$ does not reach 3/5. However, in a wide range of $r_\mathrm{D}$, it stays significantly above 2/5, which is the maximal prediction of the BJ theory (see Figures S6 and S7 in the Supporting Information for a detailed comparison of the KK and BJ theories and the respective $R(r_\mathrm{D})$ scaling plot). Overall, this behavior is consistent with the earlier mentioned theoretical expectation that the intermediate flexible chain regimes SC, saPE2, and saPE1 become narrower as $uf^{2}$ decreases. Another important observation is that, for $uf^{2} = 1/32$, the exponent consistently exceeds 3/5 (i.e., the theoretical value for the saPE2 regime) in a broad range of Debye lengths between $r_\mathrm{D} \approx 50$ and $r_\mathrm{D} \approx 400$, indicating the departure from the saPE2 regime and the onset of the development of the saPE1 regime (with the theoretical limiting slope of 1).

One more, independent theory test is the scaling of the chain size in the saPE2 regime versus the Coulomb strength parameter, $uf^{2}$, at a fixed $r_\mathrm{D}$ value. According to Table~\ref{table:1}, the theory predicts $R \sim l_\mathrm{B}^{4/15} A^{-8/15} \sim (uf^{2})^{4/15}$, and precisely these offsets, $4^{4/15}$ and $8^{4/15}$, are found between neighboring curves in the simulation data of Figure~\ref{fig:3}, as indicated by the gray bars on the left.

In summary, the performed simulations of flexible PEs fully corroborate the theoretical predictions of the KK rather than the BJ theory.

\begin{figure} [h]
\centering
\includegraphics[width=6.5in]{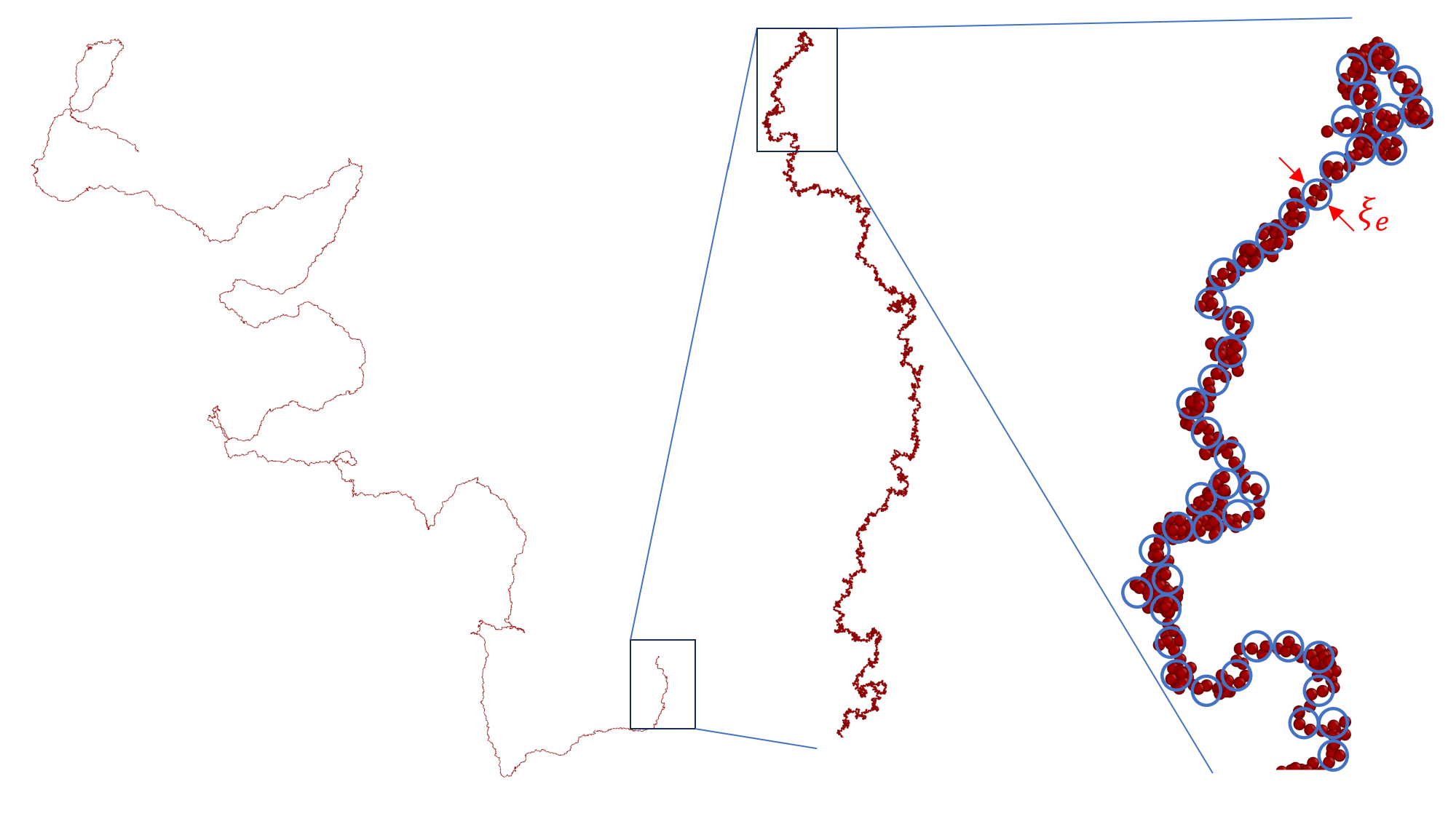}
\caption{A snapshot of a flexible polyelectrolyte chain with $N = 100000$ ($l_\mathrm{0} = 1$ and $L = 100000$) at $uf^{2} = 1/128$ and $r_\mathrm{D} = 70$. The left panel is the entire chain conformation, and the right one is zoomed in to the resolution of single monomers (beads). Blue circles at the maximum magnification qualitatively represent the Gaussian electrostatic blobs $\xi_\mathrm{e}$ that, according to the theory, should control the chain statistics at certain intermediate lengths in the saPE2 regime.}
\label{fig:4}
\end{figure}

It is now of interest to test not only the predictions of KK theory for the global chain size $R$ but also to confirm the internal conformational statistics, i.e., the underlying scaling picture of the flexible PE described first by KK~\cite{KK-1982} and elaborated in detail in the first part of the series.~\cite{EPL-part1}

\begin{figure}
\centering
\includegraphics[width=3.25in]{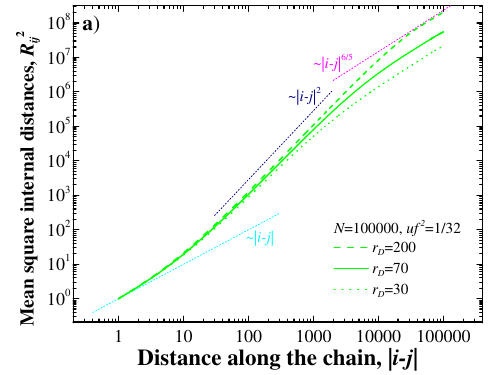} \\
\includegraphics[width=3.25in]{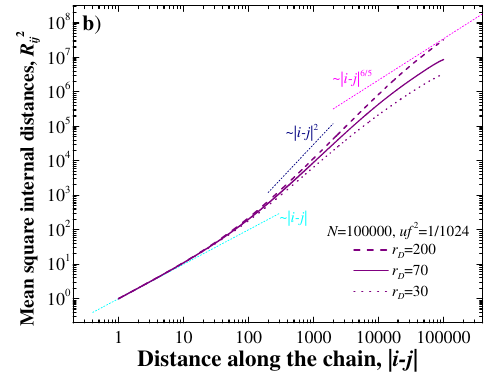}
\caption{Mean square internal distances $R_\mathrm{ij}^{2}$ for the flexible polyelectrolyte at (a) $uf^{2} = 1/32$ and (b) $uf^{2} = 1/1024$, for three different values of the Debye length $r_\mathrm{D} = 30$, 70, and 200. These parameter values correspond to the theoretical scaling regime saPE2.}
\label{fig:5}
\end{figure}

Figure~\ref{fig:4} shows a flexible PE snapshot at $uf^{2} = 1/128$ and $r_\mathrm{D} = 70$, which corresponds to the saPE2 regime. The snapshot clearly demonstrates that the chain, while being stretched globally, adopts a Gaussian electrostatic blob structure on a local scale. To study this feature in a more quantitative fashion, the mean square internal distances $R_\mathrm{ij}^{2}$ were calculated for three different values of the Debye radius at $uf^{2} = 1/32$ and $uf^{2} = 1/1024$, as demonstrated in Figure~\ref{fig:5}. Below we analyze these results by comparison with theory, see Section~\ref{sec:theory-recap} for details and Figure S1 of the Supporting Information for the theoretical internal MSD plot.

The internal chain structure at low $uf^{2}$ clearly demonstrates three regions: (i) At small contour distances $|i-j|$ along the chain, it exhibits ideal coil statistics, corroborating the Gaussian electrostatic blob structure proposed in the theory, with $R_\mathrm{ij}^{2} \sim |i-j|$; (ii) as the distance along the chain increases, the electrostatic stiffening results in the appearance of rod-like statistics, $R_\mathrm{ij}^{2} \sim |i-j|^{2}$. Note, however, that even at the highest considered $r_\mathrm{D} = 200$, the exact slope of 2 is still not reached despite having an extremely long chain, $N = 100000$. (iv) Finally, at the largest scales, the chain is a swollen coil with $R_\mathrm{ij}^{2} \sim |i-j|^{6/5}$. According to the theoretical model (see Figure S1), there should also be an intermediate region (iii) with ideal-coil statistics, $R_\mathrm{ij}^{2} \sim |i-j|$. We believe that this region, confined between two other regions with higher slopes of 2 and 6/5, is too narrow to be clearly observed even at $N = 100000$. This is similar to the situation in Figure~\ref{fig:2} for the saPE1 regime, whose theoretical slope of 1 lies between the shallower slopes of the neighboring saPE2 and sfPE regimes, 3/5 and 0, which likewise prevents it from developing fully. More broadly, this illustrates the asymptotic nature of the theoretical scaling picture.

The mean square internal distances can additionally be used to extract the Gaussian electrostatic blob size $\xi_\mathrm{e}$ to compare it to the theoretical prediction of eq.~\ref{eq:xie-ge}. To this end, we need to find the point where the $R_\mathrm{ij}^{2}$ dependence crosses over from the linear $\sim|i-j|$ to quadratic $\sim|i-j|^{2}$ behavior, and the number of monomers in the blob $g_\mathrm{e}$ can be associated with the corresponding distance along the chain. The following approach was employed: first, the linear $R_\mathrm{ij}^{2} \sim |i-j|$ scaling is drawn through the $|i-j| = 1$ and $R_\mathrm{ij}^{2} = 1$ point (corresponding to the adjacent bonded monomers); second, the quadratic $R_\mathrm{ij}^{2} \sim |i-j|^{2}$ scaling is drawn through the inflection point of the graph, that is, the point where the derivative is highest and closest to the asymptotic value of 2. The $|i-j|$ position of the intersection point provides us with an estimate for the number of monomers $g_\mathrm{e}$ in the blob, and the corresponding $R_\mathrm{int}^{2}$ at $|i-j| = g_\mathrm{e}$ is identified with the blob spatial size, $\xi_\mathrm{e}^{2}$. This procedure is shown in detail in Figure S3 of the Supporting Information.

\begin{figure}
\centering
\includegraphics[width=3.25in]{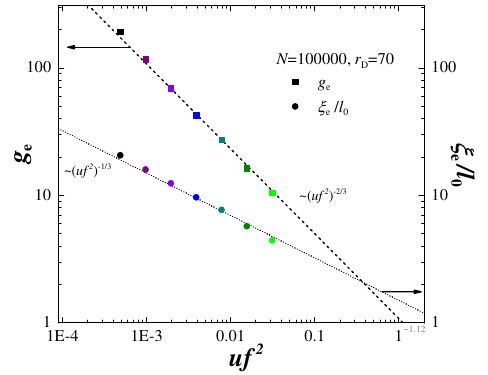}
\caption{Dependencies of the Gaussian electrostatic blob size $\xi_\mathrm{e}$ (left axis) and the number of monomers in it $g_\mathrm{e}$ (right axis) on $uf^{2}$, which were extracted from the theoretically informed analysis of the chain mean square internal distances. The straight lines with the slopes of $-1/3$ and $-2/3$ show the theoretical power laws for $\xi_\mathrm{e}$ and $g_\mathrm{e}$ given by eq.~\ref{eq:xie-ge}.}
\label{fig:6}
\end{figure}

We fixed $r_\mathrm{D} = 70$ and varied $uf^{2}$ from 1/32 to 1/2048 (having $A = 4$ and $l_\mathrm{0} = 1$ fixed for all systems and $l_\mathrm{B}$ varied), and the resulting $R_\mathrm{int}^{2}(|i-j|)$ dependencies are presented in Figure S4 of the Supporting Information. The extracted values of the spatial size of the blob $\xi_\mathrm{e}$ and the number of monomers in it $g_\mathrm{e}$ across different values of $uf^{2}$ are shown in Figure~\ref{fig:6}. One can see that the theoretical slopes of $-1/3$ and $-2/3$ are clearly reproduced, additionally corroborating the analytical scaling predictions and the proposed internal picture of the PE behind them.

Moreover, these data provide us with a numerical estimate for the flexible/semiflexible PEs crossover in terms of $uf^{2}$, which is theoretically predicted to happen at $uf^{2} \simeq 1$ but the actual numerical prefactor cannot be found within the scaling approach. Indeed, it is natural to assume that, when a Gaussian blob contains just one monomer, $g_\mathrm{e} = 1$, the chain behavior switches from flexible to semiflexible, see eq.~\ref{eq:xie-ge} and the discussion in ref.~\citenum{EPL-part1}. Figure~\ref{fig:6} shows that the $uf^{2}$ value at which this happens is approximately equal to 1.12, i.e., the flexible-semiflexible chain crossover is indeed located at $uf^{2} \approx 1$. Thus, we can speculate that the basic case of $l_\mathrm{B} = A = l_\mathrm{0} = 1$ represented in Figure~\ref{fig:2} indeed corresponds to the flexible/semiflexible crossover region on the scaling diagram of Figure~\ref{fig:1}. We should note here that, within the theoretical model (and neglecting logarithmic corrections provided in the Appendix of ref.~\citenum{EPL-part1}), the blob size $\xi_\mathrm{e}$ does not depend on $r_\mathrm{D}$, while in our simulations increasing $r_\mathrm{D}$ leads to a slight decrease in $\xi_\mathrm{e}$. In addition to that, as discussed above, region (ii) of rod-like behavior is not fully developed even for the highest studied $r_\mathrm{D} = 200$, which affects the accuracy of the localization of the transition point from local $R_\mathrm{ij}^{2} \sim |i-j|$ ideal-coil to $R_\mathrm{ij}^{2} \sim |i-j|^{2}$ rodlike statistics. Therefore, we do not attempt to localize the flexible/semiflexible crossover position with higher precision and conclude $uf^{2} \approx 1$.

Another way to obtain this result would be to formally extract the $g_\mathrm{e}$ values at $uf^{2} \gg 1$ in addition to $uf^{2} \ll 1$. Since at high $uf^{2}$ values the chain is rod-like even at the smallest scale (hence the new concept of the rodlike electrostatic blob introduced in the theory~\cite{EPL-part1} for semiflexible PEs), applying the procedure described above would always result in $g_\mathrm{e} = 1$. Finding the intersection point of the scaling laws for semiflexible chains at $uf^{2} \gg 1$, namely $g_\mathrm{e} = 1$, and flexible chains at $uf^{2} \ll 1$, namely $g_\mathrm{e} \sim (uf^{2})^{-2/3}$ shown in Figure~\ref{fig:6}, would give the same crossover point of $uf^{2} \approx 1$.

\subsection{Chain Length Effects}
\label{sec:chain-length}

\begin{figure} 
\centering
\includegraphics[width=5.0in]{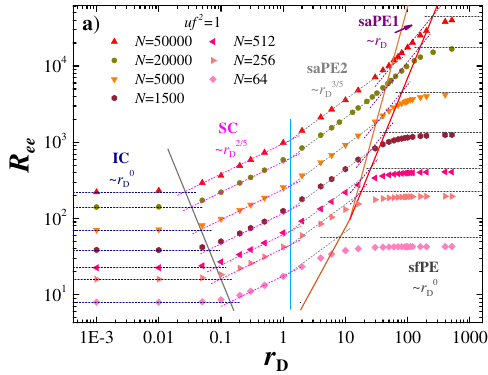} \\\
\includegraphics[width=5.0in]{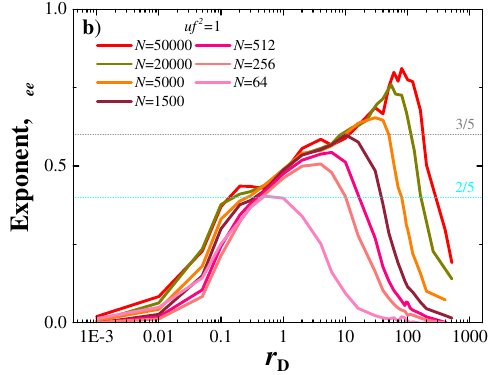}
\caption{(a) Dependencies of the polyelectrolyte end-to-end distance $R_\mathrm{ee}$ on the Debye radius $r_\mathrm{D}$ for chains of different lengths $N$ varying from 64 to 50000 at $l_\mathrm{B} = A = l_\mathrm{0} = 1$. The dashed lines indicate the positions of various scaling regimes, while the solid lines represent the crossover positions between these regimes. (b) Logarithmic derivatives of this dependence, $\alpha_\mathrm{ee} = \partial(\ln R_\mathrm{ee})/\partial(\ln r_\mathrm{D})$, corresponding to the apparent slope in the $R_\mathrm{ee} \sim r_\mathrm{D}^{\alpha_\mathrm{ee}}$ law.}
\label{fig:7}
\end{figure}

Let us now return to the flexible/semiflexible crossover region and investigate the effect of the chain length $N$ at $uf^{2} = 1$ (provided by $l_\mathrm{B} = A = l_\mathrm{0} = 1$). According to ref.~\citenum{EPL-part1} and the theoretical diagram of Figure~\ref{fig:1}, the widths of the intermediate regimes SC, saPE2, and saPE1 should increase with the chain length $N$. Figure~\ref{fig:7} shows the obtained $R_\mathrm{ee}(r_\mathrm{D})$ dependencies for $N$ varying from 64 to 50000.

One can see that increasing the chain length $N$ indeed makes the intermediate regimes wider. More importantly, Figure~\ref{fig:7}b shows that, for sufficiently short chains, the \textit{apparent} scaling exponent $\alpha_\mathrm{ee}$ can be lower than 3/5 simply because the regimes are not yet wide enough. Indeed, for $N = 64$ the maximum value is $\alpha_\mathrm{ee} = 2/5$, making it look as though the chains follow the BJ linear rather than the KK quadratic scaling for the EPL (see Figures S6 and S7 of the Supporting Information for the limiting theoretical slopes within these two theories). As the chain length increases, however, the maximum scaling exponent $\alpha_\mathrm{ee}$ grows steadily, reaching the value of 3/5 expected for the saPE2 regime at already $N = 1500$. For even longer chains, from $N = 5000$ to $50000$, the saPE1 regime begins to manifest itself, and the maximum scaling exponent grows further, tending towards the theoretically predicted value of 1 and reaching approximately 0.82 at $N = 50000$. Therefore, as already noted in Section~\ref{sec:crossover-flex-semiflex}, even longer chains are needed for $\alpha_\mathrm{ee}$ to fully reach the theoretically predicted exponent of unity, but the trend unambiguously corroborates the KK theory.

A more quantitative analysis can be carried out by applying the theoretically predicted $N$-dependences of the chain size in each regime (see Table~\ref{table:1}). For this purpose, we used the longest studied chain of $N = 50000$, and estimated the regime ranges and the crossover positions between them according to the following procedure:

\begin{enumerate}
\item In the IC regime, a horizontal line was drawn at the ideal-coil size, $R_\mathrm{ee} = l_\mathrm{0} N^{1/2} = \sqrt{50000}$, in the region of small $r_\mathrm{D}$.
\item In the SC and saPE2 regimes, lines with the theoretical slopes of 2/5 and 3/5, respectively, were drawn through the central points of each regime, located at $r_\mathrm{D} = 0.3$ and $r_\mathrm{D} = 6$. Figure~\ref{fig:7}b illustrates that these coordinates correspond to the centers of the respective plateaus in the $\alpha_\mathrm{ee}(r_\mathrm{D})$ dependence.
\item In the saPE1 regime, a line with a slope of 1 was drawn through $r_\mathrm{D} = 80$, which corresponds to the point of maximal $\alpha_\mathrm{ee}$, as seen in Figure~\ref{fig:7}b.
\item Finally, in the sfPE regime, a horizontal line was drawn at the saturated (salt-free) end-to-end size, $R_\mathrm{ee}(r_\mathrm{D} \to \infty)$.
\end{enumerate}

For the remaining chain lengths ranging from $N = 64$ to $20000$, we simply scaled the lines obtained for $N = 50000$ according to the theoretically predicted $R_\mathrm{ee}(N)$ dependence in each of the regimes, as summarized in Table~\ref{table:1}: $R \sim N^{1/2}$ for IC and saPE1, $R \sim N^{3/5}$ for SC and saPE2, and $R \sim N^{1}$ for sfPE. We emphasize that the straight regime lines for all $N$ from 64 to 20000 were therefore obtained \textit{solely} from the theory-informed re-scaling, that is, by shifting downwards the theoretical fits for $N = 50000$ data by the factor of $(N_1/N_2)^{\beta}$ with $\beta = 1/2$, $3/5$, and $1$; in other words, no direct fitting of the actual data for those shorter chains was performed. The described procedure allowed us to check the consistency of the theoretical predictions across different chain lengths.

The resulting theoretical power-law dependencies for all the regimes are shown as dashed lines in Figure~\ref{fig:7}a. For all chain lengths, we observe excellent agreement between the simulation data and the theory-informed curves shifted from the $N = 50000$ fits. As discussed above, for shorter chains, the saPE2 and especially saPE1 regimes become very narrow and do not fully develop, so that only a smooth, often not easily visually noticeable transition between neighboring regimes is observed.

It is also worth noting that, although the theoretical diagram of Figure~\ref{fig:1} predicts equal $\simeq N^{1/4}$ widths for the saPE2 and saPE1 regimes, the simulation data in Figure~\ref{fig:7}a clearly show that the saPE1 regime is noticeably narrower. For very short chains with $N < 512$, it vanishes completely. This apparent discrepancy does not actually contradict the scaling theory because the exact values of the numerical prefactors, which control the regime crossovers and widths, are not available within the scaling method; they are hidden under the $\simeq$ sign and are always neglected, i.e., are effectively set to unity. However, the performed simulations enable revealing these numerical prefactors and, in principle, making the scaling laws quantitatively exact.~\cite{dobrynin-2021} This also explains why the saPE1 regime is so difficult to detect: not only is its scaling slope of $\alpha = 1$ much larger than in the neighboring regimes (3/5 and 0 in saPE2 and sfPE), but it is also much narrower than the other intermediate regimes, SC and saPE2.

The only regime where the theoretical predictions based on $N = 50000$ data systematically overestimate the chain size for shorter chains is the essentially salt-free regime sfPE. This is due to the logarithmic corrections, which were earlier discussed in the literature~\cite{degennes-1976, DR-2005} and summarized in the Appendix of the first part of this series for all the regimes.~\cite{EPL-part1} Let us demonstrate this for the case of $N = 64$. The logarithmic correction for the sfPE regime is $\left[ \ln(N/g_\mathrm{e}) \right]^{1/3}$, and the value of $g_\mathrm{e}$ at $uf^{2} = 1$ was estimated to be $g_\mathrm{e} \approx 1$ in the previous section. Therefore, the logarithmic corrections would result in overestimating the chain size for $N = 64$ by a factor of $\left[ \ln(50000)/\ln(64) \right]^{1/3} \approx 1.38$, whereas the simulation data in Figure~\ref{fig:7}a yield 1.33, which is very close to the theoretical estimate.

The scaling lines in Figure~\ref{fig:7}a also allow us to estimate the crossover positions between regimes X and Y from their intersections. As discussed in Section~\ref{sec:theory-recap} and summarized in Table~\ref{table:2}, one can derive the $R_\mathrm{ee}^\mathrm{X/Y}(r_\mathrm{D}^\mathrm{X/Y})$ dependences, where $R_\mathrm{ee}^\mathrm{X/Y}$ and $r_\mathrm{D}^\mathrm{X/Y}$ are the crossover chain size values and Debye radius coordinates for different chain lengths $N$; these results are plotted in Figure~\ref{fig:7}a as solid lines passing through the crossover positions found for $N = 50000$. Once again, these lines perfectly connect the crossover positions across all chain lengths, demonstrating the internal consistency of the theoretical predictions. Solid lines in Figure~\ref{fig:7}a therefore provide a \textit{quantitatively accurate} diagram of scaling regimes, which can in the future serve as an excellent test for quantitative theories of the EPL.

\subsection{Orientational Correlations in Stiffened Polyelectrolyte Chains}
\label{sec:orient-corr}

\begin{figure}
\centering
\includegraphics[width=3.2in]{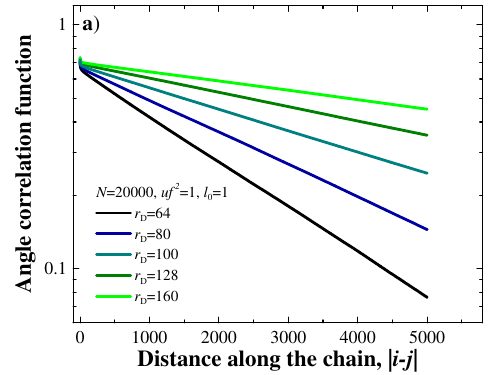}\hspace{2mm}%
\includegraphics[width=3.2in]{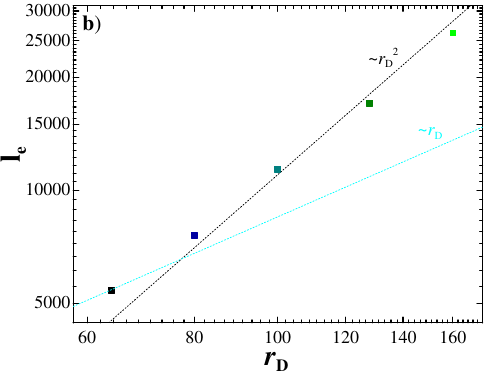}
\caption{(a) Dependence of the bond-bond angle correlation function $\left\langle \cos(\theta_\mathrm{ij}) \right\rangle$ on the distance along the chain $|i-j|$ for different $r_\mathrm{D}$ values. The dependences are cut at $|i-j| = 5000$ where reasonable averaging can still be obtained. (b) The extracted values of the slope proportional to the electrostatic persistence length $\textbf{l}_\mathrm{e}$ following the quadratic OSF/KK scaling, $\textbf{l}_\mathrm{e} \sim r_\mathrm{D}^{2}$, with the Debye radius (black dotted line); the blue dotted line shows the BJ scaling, $\textbf{l}_\mathrm{e} \sim r_\mathrm{D}^{1}$.}
\label{fig:8}
\end{figure}

Another way to explore the internal structure of a PE chain is to analyze the chain stiffness at the local level by studying the orientational (bond-bond) correlations. To this end, we calculate the bond angle correlation function for different $r_\mathrm{D}$ values at $l_\mathrm{B} = A = l_\mathrm{0} = 1$ for chains of $N = 20000$ segments ($L = N l_\mathrm{0} = 20000$), which are shown in Figure~\ref{fig:8}a. Since the angle correlation functions show how orientational correlations decay along the chain, the slopes provide access to the electrostatic persistence length $\textbf{l}_\mathrm{e}$. The single-exponential decay of these functions
\begin{equation}
    \ln \left\langle \cos(\theta_\mathrm{ij}) \right\rangle \sim - |i-j| \left( \frac{l_\mathrm{0}}{\textbf{l}_\mathrm{e}} \right)
\label{eq:corr-decay}
\end{equation}
is clearly seen, and Figure~\ref{fig:8}b summarizes the respective extracted slopes proportional to the EPL $\textbf{l}_\mathrm{e}$.

The observed exponential decay at very large distances along the chain corroborates the persistent behavior of the chain. Moreover, the recovered slopes clearly show the squared $\textbf{l}_\mathrm{e} \sim r_\mathrm{D}^{2}$ dependence, consistent with the OSF/KK theory.

\subsection{Semiflexible Polyelectrolytes}
\label{sec:semiflexible-PE}

The next region to be investigated on the diagram of Figure~\ref{fig:1} is the region of semiflexible chains, $uf^{2} \gg 1$ (see blue arrow), which is above the flexible/semiflexible crossover discussed in Section~\ref{sec:flexible-PE} and found to be located at $uf^{2} \approx 1$. Recall that the $uf^{2} = l_\mathrm{B} l_\mathrm{0}/A^{2}$ parameter quantifying the strength of electrostatic self-interactions per bare Kuhn segment can be alternatively considered as the reduced stiffness parameter: hence the flexible PE and semiflexible PE nomenclature for $uf^{2} \ll 1$ and $uf^{2} \gg 1$. Since the charge spacing $A$ was already set to the minimal possible value of $A = 1$ (each monomer is charged) when studying the case of $uf^{2} = 1$, there are two possible approaches to provide higher values of $uf^{2}$: increasing $l_\mathrm{B}$ (this section) and increasing $l_\mathrm{0}$ (Section~\ref{sec:bare-stiffness}). Given that the latter also leads to a reduction in the chain length quantified by the number of bare Kuhn segments, $N = L/l_\mathrm{0}$ (provided that the number of beads $L$ in simulations remains the same), we start with varying $l_\mathrm{B}$. 

\begin{figure}
\centering
\includegraphics[width=3.25in]{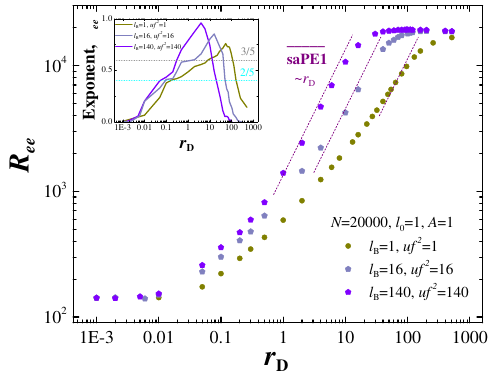}
\caption{Dependencies of the end-to-end distance $R_\mathrm{ee}$ of semiflexible polyelectrolytes on the Debye radius $r_\mathrm{D}$ at the Coulomb strength parameter $uf^{2} = 16$ and $140$ in comparison to the case of $uf^{2} = 1$. The chain length is fixed to $N = L = 20000$ and the bare stiffness is set to $l_\mathrm{0} = 1$. The inset shows the logarithmic derivatives of this dependence, $\alpha_\mathrm{ee} = \partial(\ln R_\mathrm{ee})/\partial(\ln r_\mathrm{D})$. The dashed lines show the positions of the $\overline{\text{saPE1}}$ regime, which for $uf^{2} = 16$ and $140$ were obtained from the theory-informed re-scaling.}
\label{fig:9}
\end{figure}

Figure~\ref{fig:9} shows the obtained $R_\mathrm{ee}(r_\mathrm{D})$ dependencies for substantially semiflexible chains with $uf^{2} = 16$ and $140$ in comparison to the PE with $uf^{2} = 1$ representing the crossover case. The changes in the behavior upon increasing $uf^{2}$ are in perfect agreement with the theoretical diagram of Figure~\ref{fig:1}. First, the crossovers between the regimes move to lower $r_\mathrm{D}$ values, shifting the $\alpha_\mathrm{ee}(r_\mathrm{D})$ dependence to the left. More importantly, the theory predicts that the $\overline{\text{saPE1}}$ regime becomes wider as $uf^{2}$ becomes larger provided that $N$ is fixed, and that is exactly what the simulations demonstrate. Indeed, the peak height of the logarithmic derivative $\alpha_\mathrm{ee}(r_\mathrm{D})$ grows with $uf^{2}$, reaching as high as $\alpha \approx 0.95$ at $uf^{2} = 140$. This is extremely close to the asymptotic theoretical value of $\alpha^{\overline{\text{saPE1}}} = 1$ predicted for this regime, corroborating the OSF theory of electrostatic stiffening of semiflexible PEs.~\cite{odijk-1977, SF-1977} Moreover, the positions of the $\overline{\text{saPE1}}$ regime for $uf^{2} = 16$ and $140$ were obtained from the corresponding line for $uf^{2} = 1$ (Section~\ref{sec:chain-length}, Figure~\ref{fig:7}a) using the theoretical scaling prediction for the dependence of the chain size on $l_\mathrm{B}$, $R_\mathrm{ee} \sim l_\mathrm{B}^{1/2}$ (see Table~\ref{table:1}). The simulation data are in excellent agreement with the predicted behavior, further corroborating the consistency of the scaling theory presented in the first part of this series.~\cite{EPL-part1}

\subsection{Effect of the Bare Stiffness $l_\mathrm{0}$}
\label{sec:bare-stiffness}

\begin{figure}
\centering
\includegraphics[width=3.25in]{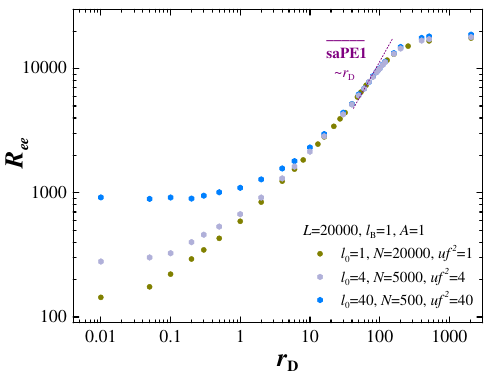}
\caption{Dependences of the end-to-end distance $R_\mathrm{ee}$ of semiflexible polyelectrolytes on the Debye radius $r_\mathrm{D}$ for chains with different bare stiffness $l_\mathrm{0} = 4$ and $40$ in comparison to the chain with $l_\mathrm{0} = 1$ corresponding to the flexible/semiflexible crossover. Chains comprise an equal number of chemical monomers, $L = 20000$.}
\label{fig:10}
\end{figure}

A separate question, however, remains: how does changing solely the bare stiffness, $l_\mathrm{0}$, affect the system behavior? To investigate it, we studied the system with $l_\mathrm{B} = 1$, $A = 1$, contour length $L = 20000$, and three different bare Kuhn segment values $l_\mathrm{0}$ of 1, 4, and 40 (the corresponding $uf^{2}$ values are also 1, 4, and 40), thus once again exploring the region of semiflexible chains. Note that these parameters correspond to chains with different numbers of Kuhn segments $N$, namely $N = 20000$, 5000, and 500, respectively. Figure~\ref{fig:10} provides the obtained dependencies $R_\mathrm{ee}(r_\mathrm{D})$ of the chain size on the Debye screening radius.

Upon increasing the bare stiffness $l_\mathrm{0}$, the chain behavior at low $r_\mathrm{D}$ changes (namely, the chain size increases), while at high $r_\mathrm{D}$ the size remains universally the same. Note that the abscissa coordinate $r_\mathrm{D}$ is different from that of the theoretical diagram in Figure~\ref{fig:1}, $r_\mathrm{D}/l_\mathrm{0}$, because $l_\mathrm{0}$ also varies across the curves. The different behavior at low $r_\mathrm{D}$ can be easily understood by remembering that at strong salt screening, the effect of electrostatic stiffening is weak, and the chain conformations are primarily governed by the bare stiffness $l_\mathrm{0}$, so the chain size grows with increasing $l_\mathrm{0}$. At high values of $r_\mathrm{D}$, however, when the electrostatic persistence length becomes much larger than the bare stiffness, $\textbf{l}_\mathrm{e} \gg l_\mathrm{0}$, the chain size becomes controlled by the former, so there is no change in the chain size upon changing $l_\mathrm{0}$. It is important to note that this behavior cannot be easily grasped from the theoretical diagram in Figure~\ref{fig:1} because increasing $l_\mathrm{0}$ alone, while \textit{not} maintaining the fixed $N = L/l_\mathrm{0}$ value, does not correspond to moving vertically up on the same diagram. Indeed, changing $l_\mathrm{0}$ at fixed $L$ rather than fixed $N$ rescales the entire diagram, including the regime widths and the coordinates of points where multiple regimes meet and terminate, both of which depend on $N$. This is, however, completely consistent with the scaling laws summarized in Table~\ref{table:1}, which predict no dependence of the chain size on $l_{0}$ in regimes $\overline{\text{saPE1}}$ and $\overline{\text{saPE2}}$.

More broadly, one can say that polymers with higher bare stiffness are less sensitive to variations in the Debye radius $r_\mathrm{D}$ (or, equivalently, in the salt concentration) than intrinsically flexible ones: their size varies less between the limiting cases of fully screened and unscreened interactions. This trend is consistent with the experimentally observed behavior of intrinsically stiff DNA and intrinsically flexible PSS chains over a range of salt concentrations.~\cite{samghabadi-2025}

A simple scaling analysis illustrating why the curves collapse on top of each other at high $r_\mathrm{D}$ values (in the $\overline{\text{saPE2}}$, $\overline{\text{saPE1}}$, and $\overline{\text{sfPE}}$ regimes) can be performed. Using eqs.\ 57 and 60 of the theoretical part of the series,~\cite{EPL-part1} the left and right boundaries of the $\overline{\text{saPE1}}$ regime can be estimated as
\begin{equation}
    r_\mathrm{D}^{\overline{\text{saPE2}}/\overline{\text{saPE1}}} \simeq \left( \frac{L A^{6}}{l_\mathrm{B}^{3}} \right)^{1/4}
\label{eq:rD-saPE2bar-saPE1bar}
\end{equation}
\begin{equation}
    r_\mathrm{D}^{\overline{\text{saPE1}}/\overline{\text{sfPE}}} \simeq \left( \frac{L A^{2}}{l_\mathrm{B}} \right)^{1/4}
\label{eq:rD-saPE1bar-sfPEbar}
\end{equation}
They are independent of the chain bare stiffness $l_\mathrm{0}$ (provided that $uf^{2} \ll N$, see Figure~\ref{fig:1}), and so is the total $r_\mathrm{D}$-width of the regime $\overline{\text{saPE1}}$, equal to
\begin{equation}
    \frac{r_\mathrm{D}^{\overline{\text{saPE1}}/\overline{\text{sfPE}}}}{r_\mathrm{D}^{\overline{\text{saPE2}}/\overline{\text{saPE1}}}} \simeq (uf^{2})^{1/4} N^{1/4} = \left( \frac{L}{l_\mathrm{0}} \right)^{1/4} \left( \frac{l_\mathrm{0} l_\mathrm{B}}{A^{2}} \right)^{1/4} = \left( \frac{L l_\mathrm{B}}{A^{2}} \right)^{1/4}
\label{eq:width-saPE1bar-region}
\end{equation}
This clarifies why the position and width of the regime $\overline{\text{saPE1}}$ remain the same across different $l_\mathrm{0}$, providing the collapse of the different $R_\mathrm{ee}(r_\mathrm{D})$ curves in this region of $r_\mathrm{D}$.

At the same time, the curves start departing from each other at low $r_\mathrm{D}$ values because the left boundary of the $\overline{\text{saPE2}}$ regime, corresponding to the OH/$\overline{\text{saPE2}}$ crossover and given by
\begin{equation}
    r_\mathrm{D}^{\mathrm{OH}/\overline{\text{saPE2}}} \simeq \left( \frac{A^{2} l_\mathrm{0}}{l_\mathrm{B}} \right)^{1/2},
\label{eq:rD-OH-saPE2bar}
\end{equation}
is a function of the bare stiffness $l_\mathrm{0}$, as follows from eq.\ 64 of the theoretical part.~\cite{EPL-part1} Since the $\overline{\text{saPE2}}/\overline{\text{saPE1}}$ crossover is $l_\mathrm{0}$-independent (eq.~\ref{eq:rD-saPE2bar-saPE1bar}), the width of the regime $\overline{\text{saPE2}}$,
\begin{equation}
    \frac{r_\mathrm{D}^{\overline{\text{saPE2}}/\overline{\text{saPE1}}}}{r_\mathrm{D}^{\mathrm{OH}/\overline{\text{saPE2}}}} \simeq (uf^{2})^{-1/4} N^{1/4} = \left( \frac{L A^{2}}{l_\mathrm{B} l_\mathrm{0}^{2}} \right)^{1/4}
\label{eq:width-saPE2bar-region}
\end{equation}
decreases with $l_\mathrm{0}$, as clearly seen in Figure~\ref{fig:10}. Finally, we note that slight differences in chain size are observed in the $\overline{\text{sfPE}}$ regime, which can be attributed to the different chain finite extensibility at different values of bare stiffness $l_\mathrm{0}$.

The final regime from the theoretical diagram (Figure~\ref{fig:1}) that we have not yet detected is the OH regime with $R \sim r_\mathrm{D}^{1/5}$; it does not clearly appear as a separate entity in either Figure~\ref{fig:9} or Figure~\ref{fig:10}. We believe the reason for this is threefold. (i) First of all, for $uf^{2} \ll N^{1/2}$ primarily available in simulations, the OH regime is confined between the SC and $\overline{\text{saPE2}}$ regimes, both with higher slopes of 2/5 and 3/5 versus $r_\mathrm{D}$, as specified in Table~\ref{table:1}. This naturally suppresses the development of this asymptotic regime at low chain lengths $N$. (ii) Second, as with the saPE1 regime discussed in Section~\ref{sec:chain-length}, the OH regime width is most likely lower than predicted, owing to the prefactors neglected (and in principle unavailable) within the scaling theory. Large values of $N$ would therefore be required to observe it. (iii) This is where the third factor comes into play: as one can see from the theoretical diagram of Figure~\ref{fig:1}, the OH regime lies at rather low $r_\mathrm{D}/l_\mathrm{0}$ values. As noted in Section~\ref{sec:param-limits} discussing the parameter limitations, the OH regime requires $A \ll r_\mathrm{D}$ to ensure that the Debye blobs of adjacent charges overlap and form a continuous cylindrical shell around the chain. Since the minimal available value of the charge spacing is $A = 1$ in our simulation model with discrete beads, the OH regime therefore requires $r_\mathrm{D} \gg 1$ and cannot develop for $r_\mathrm{D} < 1$; instead, it is replaced by the SC regime, characterized by simple excluded-volume interactions arising from weak non-cooperative electrostatic repulsions of individual charged monomers (see Section V in ref.~\citenum{EPL-part1}). This situation corresponds to increasing $uf^{2}$ by raising $l_\mathrm{B}$, as was done in Section~\ref{sec:semiflexible-PE}. To detect the OH regime, one must instead increase $uf^{2}$ by raising $l_\mathrm{0}$ alone and keep $l_\mathrm{B} = 1$ and $A = 1$ fixed. Indeed, if we consider $uf^{2} \simeq N^{1/2}$, that is, bare stiffness as high as $l_\mathrm{0} \simeq N^{1/2}$ -- the $uf^{2}$ value where the OH regime is predicted to be the widest (see Figure~\ref{fig:1}) -- the left boundary of the regime should appear at $r_\mathrm{D}/l_\mathrm{0} \simeq N^{-1/2}$, or equivalently, $r_\mathrm{D} \simeq 1$. This, however, translates into a chain length of $L = N l_\mathrm{0} \simeq N^{3/2}$, so that studying PEs with $N = 10^{4}$ bare Kuhn segments would require having $L = 10^{6}$ chemical monomers (beads). Currently, this is completely computationally unfeasible.

\subsection{Universality of the Theoretical Coordinates}
\label{sec:universality}

Finally, we would like to perform a universality analysis of the diagram. Indeed, the analytical results of Figure~\ref{fig:1} predict that the PE behavior should be universal in coordinates $uf^{2}$ versus $r_\mathrm{D}/l_\mathrm{0}$. We used that prediction in the previous sections, but is it in fact justified? To test that, we fixed $A = 1$ and used three values of $l_\mathrm{0}$ equal to 1, 4, and 12, simultaneously decreasing the value of $l_\mathrm{B}$ so that the value of $uf^{2} = l_\mathrm{0} l_\mathrm{B}/A^{2}$ is fixed at 1. Increasing $l_\mathrm{0}$ leads to a reduction of the chain length $N$ in theory: it is defined as $N = L/l_\mathrm{0}$, with $L$ being the chain contour length, because $N$ represents the number of bare Kuhn segments rather than chemical monomers (simulation beads) in the chain. To take that into account and provide the fixed $N = 1500$ value, the contour length of the chain was adjusted accordingly to $L = 1500$, $6000$, and $18000$ for the respective values of $l_\mathrm{0} = 1$, 4, and 12.

\begin{figure}
\centering
\includegraphics[width=3.25in]{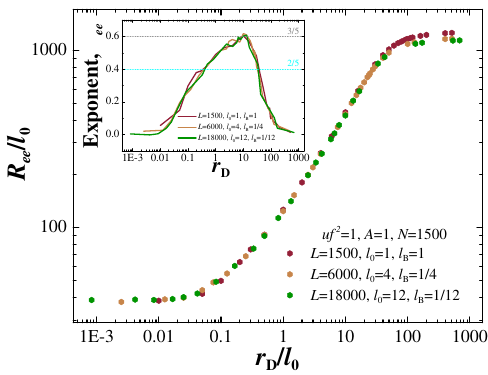}
\caption{Dependencies of the reduced polyelectrolyte size $R_\mathrm{ee}/l_\mathrm{0}$ on the reduced Debye length $r_\mathrm{D}/l_\mathrm{0}$ for three different values of the bare chain stiffness $l_\mathrm{0}$. The chain lengths $L$ were adjusted to keep the number of bare chain segments $N$ constant. The inset shows the corresponding logarithmic derivatives $\alpha_\mathrm{ee} = \partial(\ln R_\mathrm{ee})/\partial(\ln r_\mathrm{D})$ of the main dependence.}
\label{fig:11}
\end{figure}

Figure~\ref{fig:11} shows the obtained dependencies of $R_\mathrm{ee}/l_\mathrm{0}$ on $r_\mathrm{D}/l_\mathrm{0}$. Here, the chain size $R_\mathrm{ee}$ was normalized by $l_\mathrm{0}$ (being the only length scale if one operates with the dimensionless theoretical parameters $u$, $f$, $r_\mathrm{D}/l_\mathrm{0}$, and $N$, as in the first paper of the series~\cite{EPL-part1}) to account for the larger size of the chain Kuhn segments.

The correspondence between all three curves is perfect, and all the data fall on a single universal master curve. This is especially apparent when investigating the exponent $\alpha$ dependence shown in the inset in Figure~\ref{fig:11}. The only difference between the systems is a slightly different chain size $R_\mathrm{ee}/l_\mathrm{0}$ at very large $r_\mathrm{D}/l_\mathrm{0}$, which is especially pronounced in the sfPE regime: indeed, as the number of monomers goes up, the chain size goes slightly down. This behavior can be attributed to the different extensibilities of freely-jointed (i.e., for low $l_\mathrm{0}$ values) and worm-like chains (i.e., for high $l_\mathrm{0}$ values).~\cite{DR-2010}

Therefore, the collapse of the simulation data onto the master curve once again corroborates the theoretical predictions suggesting the universality in coordinates of the Kuhn segment's electrostatic strength $uf^{2}$ versus the reduced Debye radius $r_\mathrm{D}/l_\mathrm{0}$.

\section{Comparison to Experiment}
\label{sec:experiment}

The \textit{asymptotic} nature of the scaling laws revealed by our simulations and summarized in Figure~\ref{fig:7} explains why different \textit{apparent} slopes can be observed in experiments performed on PEs of different lengths. This, in turn, clarifies a broad body of experimental literature reporting support for either BJ or KK scaling. We emphasize that observing an apparent slope of 2/5 for relatively short chains should be regarded as neither proof of BJ scaling nor a contradiction of KK scaling, as illustrated by the curves for $N = 64$ and even for $N = 256$. By contrast, experimentally observing a slope that noticeably exceeds 2/5 for long chains does disprove the BJ picture, as the latter cannot account for any scaling slope higher than this value (see Figures S6-S7 in the Supporting Information for details).

Therefore, in order to test the limiting scaling law for the EPL, one should study sufficiently (in practice, very) long PEs, whose behavior comes closest to the limiting one. In this regard, results for the chain gyration radius (obtained from static light scattering, SLS) are of greater interest and have a much higher potential to approach the limiting scaling laws than those for the hydrodynamic radius (obtained from dynamic light scattering, DLS), as clearly demonstrated in Figure~\ref{fig:2} and elaborated in the accompanying discussion of Section~\ref{sec:crossover-flex-semiflex}.

\begin{figure}
\centering
\includegraphics[width=3.25in]{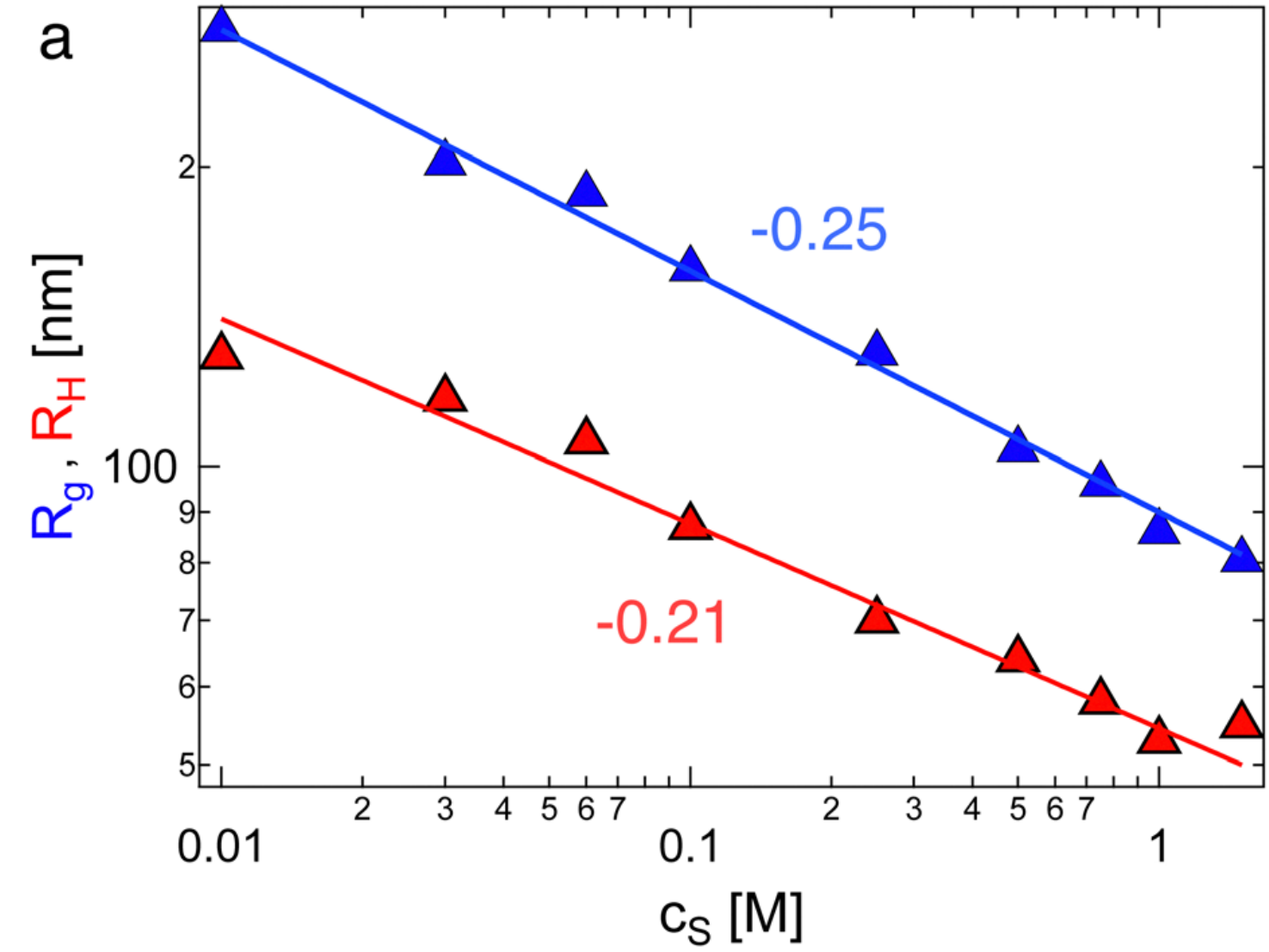}
\caption{Experimental dependence of the gyration radius (blue) and hydrodynamic radius (red) of sodium polyacrylate on the concentration of the added NaCl salt. The chain degree of polymerization is $DP = 35000$, which corresponds to the theoretical number of Kuhn segments $N \approx 6000$. Lines are the best power-law fits to the data: $R_\mathrm{g} \sim c_\mathrm{s}^{-0.25}$ (blue line) and $R_\mathrm{h} \sim c_\mathrm{s}^{-0.21}$ (red line). Adapted with permission from refs.~\citenum{schweins-2003} and~\citenum{review-dilute-PE}.}
\label{fig:12}
\end{figure}

The results of ref.~\citenum{schweins-2003}, from which the relevant scaling slope was extracted by L\'{o}pez et al.\ in ref.~\citenum{review-dilute-PE} and which are reproduced in Figure~\ref{fig:12}, provide an excellent illustration of the above. The sodium polyacrylate studied in ref.~\citenum{schweins-2003} comprised $N \approx 6000$ statistical segments, and the measured dependencies of its gyration radius and hydrodynamic radius on the concentration of added NaCl are best fit by $R_\mathrm{g} \sim c_\mathrm{s}^{-0.25}$ (red line) and $R_\mathrm{h} \sim c_\mathrm{s}^{-0.21}$ (blue line), respectively. This corresponds to $R_\mathrm{g} \sim r_\mathrm{D}^{0.5}$ and $R_\mathrm{h} \sim r_\mathrm{D}^{0.42}$ dependencies on the Debye radius $r_\mathrm{D}$. First, the exponent for the hydrodynamic radius is substantially lower, as anticipated from our simulations (cf.\ Figure~\ref{fig:2}); it only marginally exceeds the BJ value of 2/5, so on their own these $R_\mathrm{h}$ data are insufficient to corroborate either BJ or KK scaling. Second, the 0.50 exponent for the gyration radius substantially (statistically significantly) exceeds the $2/5 = 0.4$ value, thereby serving as strong experimental evidence against the BJ predictions and in favor of the KK result.

Solving the problem discussed here would benefit from further, more meticulous experimental verification, so that the long-standing contradiction concerning the EPL of flexible PEs can be resolved completely -- not only in theory and simulations, but also unambiguously in experiment. We suggest testing a series of PEs of different lengths and, in the spirit of Figure~\ref{fig:2}, examining how the apparent slope in the dependence of the chain size $R$ on the salt concentration $c_\mathrm{s}$ (equivalently, the Debye radius $r_\mathrm{D}$) evolves with the chain length $N$. In doing so, measuring the gyration radius $R_\mathrm{g}$ by SLS should be strongly preferred over studying the hydrodynamic radius $R_\mathrm{h}$ by DLS, because the former better reflects the global rather than the local chain statistics.

\begin{figure} 
\centering
\includegraphics[width=3.25in]{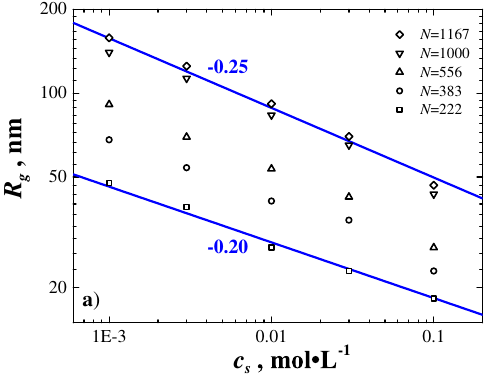} \\
\includegraphics[width=3.25in]{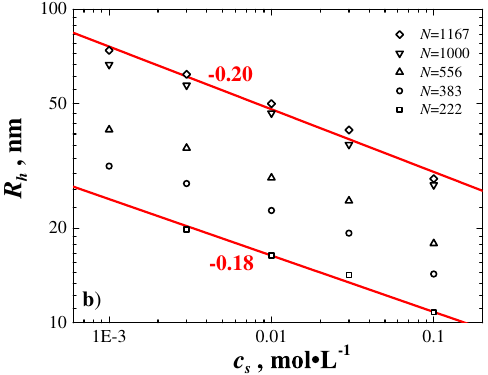}
\caption{Experimental dependence of the (a) gyration radius $R_\mathrm{g}$ and (b) hydrodynamic radius $R_\mathrm{h}$ of the bromide salts of benzyl-quaternized poly(2-vinylpyridine), Bz-P2VP Br, on the concentration of the added NaBr salt. Different curves correspond to chain contour lengths of $L_\mathrm{w} = 400$, $690$, $1000$, $1800$, and $2100$ nm. Estimating the Kuhn segment length as $l_\mathrm{0} = 1.8$ nm leads to chain lengths ranging from $N = 222$ to $1167$, as shown in the legend. Lines are the power-law fits to the data corresponding to the shortest chains, $R_\mathrm{g} \sim c_\mathrm{s}^{-0.20}$ and $R_\mathrm{h} \sim c_\mathrm{s}^{-0.18}$, and the longest chains, $R_\mathrm{g} \sim c_\mathrm{s}^{-0.25}$ and $R_\mathrm{h} \sim c_\mathrm{s}^{-0.20}$. Adapted with permission from ref.~\citenum{beer-1997}.}
\label{fig:13}
\end{figure}

A study of this type has been carried out on the bromide salts of benzyl-quaternized poly(2-vinylpyridine), Bz-P2VP Br, for chains with contour lengths ranging from $L_\mathrm{w} = 400$ to $2100$ nm.~\cite{beer-1997} Roughly assuming the bare Kuhn segment of hydrophobically modified PVP to be equal to (or larger than) that of ordinary PVP, $l_\mathrm{0} = 1.8$ nm, one can estimate (as an upper bound) that the number of Kuhn segments varies between $N \approx 222$ and $N \approx 1167$. The dependence of the chain gyration radius $R_\mathrm{g}$ on the concentration of added 1:1 NaBr salt reported in ref.~\citenum{beer-1997}, together with the corresponding scaling fits $R_\mathrm{g} \sim c_\mathrm{s}^{-\alpha/2}$, is shown in Figure~\ref{fig:13}a. The resulting slope (independent of the exact $l_\mathrm{0}$ value) increases from $-\alpha_\mathrm{g}/2 = -0.20$ for the shortest chains to $-\alpha_\mathrm{g}/2 = -0.25$ for the longest studied.

This is in reasonable qualitative agreement with the simulation results of Figure~\ref{fig:7}, where short chains display an apparent slope of $\alpha_\mathrm{ee} = 0.4$, while PEs with $N = 512$ exhibit a maximum slope of $\alpha_\mathrm{ee} = 0.5$. The difference in chain lengths between experiment and simulations should be primarily attributed to $\alpha_\mathrm{ee} \geq \alpha_\mathrm{g}$, as shown in Figure S5 of the Supporting Information, which presents simulation $\alpha_\mathrm{g}$ dependencies across different chain lengths and does provide a better \textit{quantitative} agreement with the experiment: for $N=1500$, the maximum value is indeed $\alpha_\mathrm{g} \approx 0.5$. 

In any case, the experimental slope $\alpha_\mathrm{g}$ increasing up to 0.5 is clearly higher than the maximal BJ value of 0.4, but not yet reaching the value of 0.6 characteristic of the fully developed KK saPE2 regime. Unfortunately, PEs with chain lengths $N > 5000$, which according to Figures~\ref{fig:7} and S5 are expected to exhibit a slope of $\alpha_\mathrm{g} = 0.6$ that would unambiguously corroborate the KK scaling, were not studied in ref.~\citenum{beer-1997}. We hope that the present work will stimulate experimental studies exploring such high $N$ values.

Finally, we note that the hydrodynamic radius data studied in ref.~\citenum{beer-1997} are shown in Figure~\ref{fig:13}b. They systematically exhibit lower apparent slope values, $\alpha_\mathrm{h} < \alpha_\mathrm{g}$, confirming the desired focus on the gyration rather than hydrodynamic radius in experimental studies.

\section{Conclusions}
\label{sec:conclusions}

The present and the accompanying paper~\cite{EPL-part1} represent a comprehensive simulation and theoretical study of the long-lasting problem of the electrostatic persistence length (EPL). Our key findings, which (we hope) resolve most of the contradictions, are as follows:

(i) A comprehensive theoretical diagram of limiting scaling regimes has been constructed, which reveals deep similarities in the electrostatic stiffening of semiflexible and flexible polyelectrolytes (see Figure~\ref{fig:1}). Flexible chains exhibit stiffening in accordance with the Khokhlov-Khachaturian (KK) scaling, with the electrostatic persistence length increasing quadratically with the Debye radius, $\textbf{l}_\mathrm{KK} \simeq r_\mathrm{D}^{2}/\xi_\mathrm{e}$. This is completely analogous to the Odijk-Skolnick-Fixman (OSF) scaling for semiflexible chains, $\textbf{l}_\mathrm{OSF} \simeq r_\mathrm{D}^{2}/\xi_\mathrm{e}^{|}$. The only difference is the definition of the electrostatic blob: a locally Gaussian blob $\xi_\mathrm{e}$ for flexible polyelectrolytes versus a locally rodlike blob $\xi_\mathrm{e}^{|}$ for semiflexible counterparts.~\cite{EPL-part1}

(ii) The asymptotic validity of the OSF/KK stiffening is corroborated by our coarse-grained simulations of model ideal polyelectrolyte chains with implicit salt, i.e., polymers whose charged monomers interact via a screened Coulomb potential. In the limit of very long flexible chains comprising $N \simeq 10^{4} - 10^{5}$ bare Kuhn segments, all theoretically predicted scaling regimes are virtually reproduced. Scaling for both the total chain size (end-to-end distance $R_\mathrm{ee}$) and the decay of the orientational correlations at large length scales is consistent with the KK theory, confirming the $\textbf{l}_\mathrm{KK} \sim r_\mathrm{D}^{2}$ scaling. Moreover, for flexible PEs with sufficiently low linear charge density, the chain is shown to consist of Gaussian electrostatic blobs of size $\xi_\mathrm{e}$, whose dependence on the electrostatic strength parameter $uf^2$ follows the predicted power laws.

(iii) An important insight obtained from the simulations is that the saPE2 and saPE1 regimes, unambiguously indicating and unique to the KK scaling picture of the electrostatic stiffening, are well developed and can only be clearly observed for very long and extremely long polyelectrolytes, respectively. For instance, at the flexible/semiflexible crossover with $uf^{2} = l_\mathrm{0} l_\mathrm{B}/A^{2} \simeq 1$, these values are of the order of $N_\mathrm{saPE2} \simeq 5 \times 10^{3}$ and $N_\mathrm{saPE1} > 5 \times 10^{4}$. Only above these lengths does the apparent scaling slope $\alpha$, calculated as the logarithmic derivative of the dependence of the chain end-to-end distance on the Debye radius, $R_\mathrm{ee} \sim r_\mathrm{D}^{\alpha}$, reaches $\alpha = 3/5$ (development of saPE2) and starts to substantially exceed this value by continuously albeit very slowly approaching $\alpha = 1$ ( development of saPE1). Such high values of $\alpha$ observed in the performed Monte Carlo simulations completely rule out the BJ alternative with the linear scaling of the electrostatic persistence length, $\textbf{l}_\mathrm{BJ} \sim r_\mathrm{D}$, because within that theory $\alpha$ can never exceed 2/5.

(iv) These conclusions should be taken into account when interpreting experimental results. Namely, observing (for the gyration radius) a scaling of $R_\mathrm{g} \sim c_\mathrm{s}^{-1/5}$, which is seemingly consistent with the BJ theory, \textit{does not corroborate} its validity and only indicates that the polyelectrolytes studied are not sufficiently long to truly distinguish between the BJ and KK predictions. In contrast, detecting experimental scaling $R_\mathrm{g} \sim c_\mathrm{s}^{-\alpha/2}$ with a robust $\alpha > 2/5$ is direct evidence against the BJ theory and in favor of the KK theory. Experimental values of $\alpha \approx 0.5$ have been reported by several independent groups studying very long polyelectrolytes of different chemistry,~\cite{schweins-2003, review-dilute-PE, beer-1997} with the slope $\alpha$ continuously increasing from $\alpha \approx 2/5 = 0.4$ for short chains to $\alpha \approx 0.5$ for the longest chains,~\cite{beer-1997} in almost quantitative agreement with our simulations (see Figure S5).

(v) Additionally, our simulations reveal that different measures of the chain size begin to exhibit the asymptotic theoretical exponents at different chain lengths $N$: the required $N$ is smallest for the end-to-end distance $R_\mathrm{ee}$, larger for the gyration radius $R_\mathrm{g}$, and even larger for the hydrodynamic radius $R_\mathrm{h}$, as shown in Figure~\ref{fig:2} and Figure S5. The discovered hierarchy of the respective \textit{apparent} exponents at a fixed $N$, $\alpha_\mathrm{ee} \geq \alpha_\mathrm{g} \geq \alpha_\mathrm{h}$, is consistent with experiments, where higher slopes are normally observed for $R_\mathrm{g}$ compared to $R_\mathrm{h}$ (see Figures~\ref{fig:12} and~\ref{fig:13}). The exponent for the latter did not exceed the threshold value of $\alpha_\mathrm{h}/2 = 0.2$ in the available experimental studies, but in view of the systematically reported $\alpha_\mathrm{g}/2 > 0.2$, the results for the hydrodynamic radius can be considered neither a proof of the BJ scaling nor a contradiction of the KK theory. Therefore, in future experimental studies, focusing on $R_\mathrm{g}$ rather than $R_\mathrm{h}$ appears more promising.

In conclusion, the first part of our work derives the theoretical asymptotic laws.~\cite{EPL-part1} The present part not only corroborates them in simulations as the limiting result for extremely long chains, but also provides estimates of how long PE chains should be to observe these laws in real experiments. Informed by theory and simulations, our analysis of the available experimental data shows that these data rule out the linear BJ scaling and indirectly support the KK model of electrostatic stiffening.

We hope that the present study resolves most of the experimental contradictions surrounding the electrostatic persistence length problem, which stem primarily from identifying the apparent slopes measured in experiment with the true asymptotic exponents. The latter are given by the OSF/KK theory, with the electrostatic persistence length quadratic in the Debye radius both for flexible and semiflexible chains, $\textbf{l}_\mathrm{e} \sim r_\mathrm{D}^{2}$.

Our results also explain the practical success of the scaling theory of semidilute polyelectrolyte solutions developed by Dobrynin, Colby, and Rubinstein,~\cite{DCR-1995, DR-2005, D-2020} which is based on the linear BJ scaling for the EPL. In real experiments, the polymer fragment whose size equals the solution correlation length (mesh size) -- the fragment that undergoes electrostatic stiffening before encountering other chains -- rarely exceeds one hundred (or at most a few hundred) monomers, so the BJ result appears as the \textit{effective} scaling. Only the study of very long PE chains at very low concentrations, yet still above the overlap concentration, is likely to reveal significant deviations from it and the onset of the crossover to the asymptotic KK power law, similar to the behavior of dilute solutions shown in Figure~\ref{fig:13}.

\section*{Acknowledgment}
A.M.R. and A.A.G. gratefully acknowledge Michael Rubinstein for detailed discussions. They are also thankful to Andrey Dobrynin for useful conversations.


\begin{thebibliography}{99}

\bibitem{EPL-part1}
Rumyantsev, A. M.; Gavrilov, A. A.; Johner, A.
Electrostatic persistence length revisited. I. Theory.
{\it Submitted.}

\bibitem{KK-1982}
Khokhlov, A. R.; Khachaturian, K. A.  
On the theory of weakly charged polyelectrolytes. 
{\it Polymer} {\bf 1982}, 23, 1742--1750.

\bibitem{odijk-1977}
Odijk, T.  
Polyelectrolytes near the rod limit. 
{\it J. Polym. Sci. Polym. Phys. Ed.} {\bf 1977}, 15, 477--483.

\bibitem{SF-1977}
Skolnick, J.; Fixman, M.  
Electrostatic persistence length of a wormlike polyelectrolyte.
{\it Macromolecules} {\bf 1977}, 10, 944--948.

\bibitem{barrat-1993}
Barrat, J. L.; Boyer, D.
Numerical study of a charged bead-spring chain.
{\it J. Phys. II (France)} {\bf 1993}, 3, 343--356.

\bibitem{kremer-1997}
Micka, U.; Kremer, K.
Persistence length of weakly charged polyelectrolytes with variable intrinsic stiffness.
{\it Europhys. Lett.} {\bf 1997}, 38, 279--284.

\bibitem{ullner-2002}
Ullner, M.; Woodward, C. E.
Orientational correlation function and persistence lengths of flexible polyelectrolytes.
{\it Macromolecules} {\bf 2002}, 35, 1437--1445.

\bibitem{everaers-2002}
Everaers, R.; Milchev, A.; Yamakov, V.
The electrostatic persistence length of polymers beyond the OSF limit.
{\it Eur. Phys. J. E} {\bf 2002}, 8, 3--14.

\bibitem{shklovskii-2002}
Nguyen, T. T.; Shklovskii, B. I.
Persistence length of a polyelectrolyte in salty water: Monte Carlo study.
{\it Phys. Rev. E} {\bf 2002}, 66, 021801.

\bibitem{dobrynin-2009}
Gubarev, A.; Carrillo, J.-M. Y.; Dobrynin, A. V.  
Scale-dependent electrostatic stiffening in biopolymers. 
{\it Macromolecules} {\bf 2009}, 42, 5851--5860.

\bibitem{DC-2009}
Dobrynin, A. V.; Carrillo, J.-M. Y. 
Swelling of biological and semiflexible polyelectrolyte. 
{\it J. Phys.: Condens. Matter} {\bf 2009}, 21, 424112. 

\bibitem{prochazka-2012}
Ba\v{c}ov\'{a}, P.; Ko\v{s}ovan, P.; Uhl\'{i}k, F.; Kuldov\'{a}, J.; Limpouchov\'{a}, Z.; Proch\'{a}zka, K.
Double-exponential decay of orientational correlations in semiflexible polyelectrolytes.
{\it Eur. Phys. J. E} {\bf 2012}, 35, 53.

\bibitem{buehler-2012}
Cranford, S. W.; Buehler, M. J.
Variation of weak polyelectrolyte persistence length through an electrostatic contour length.
{\it Macromolecules} {\bf 2012}, 45, 8067--8082.

\bibitem{stevens-2018}
Stevens, M. J.; Berezney, J. P.; Saleh, O. A.
The effect of chain stiffness and salt on the elastic response of a polyelectrolyte.
{\it J. Chem. Phys.} {\bf 2018}, 149, 163328.

\bibitem{BJ-1993}
Barrat, J.-L.; Joanny, J.-F.   
Persistence length of polyelectrolyte chains. 
{\it Europhys. Lett.} {\bf 1993}, 24, 333--338.

\bibitem{carrillo-2013}
Carrillo, J.-M. Y.; MacKintosh, F. C.; Dobrynin, A. V.
Nonlinear elasticity: From single chain to networks and gels.
{\it Macromolecules} {\bf 2013}, 46, 3679--3692.

\bibitem{madras-1988}
Madras, N.; Sokal, A. D.
The pivot algorithm: A highly efficient Monte Carlo method for the self-avoiding walk.
{\it J. Stat. Phys.} {\bf 1988}, 50, 109--186.

\bibitem{GK-book}
Grosberg, A. Yu.; Khokhlov, A. R.
{\it Statistical Physics of Macromolecules};
AIP Press: New York, 1994.

\bibitem{dobrynin-2021}
Dobrynin, A. V.; Jacobs, M.; Sayko, R.
Scaling of polymer solutions as a quantitative tool.
{\it Macromolecules} {\bf 2021}, 54, 2288--2295.

\bibitem{degennes-1976}
de Gennes, P.-G.; Pincus, P.; Velasco, R. M.; Brochard, F.
Remarks on polyelectrolyte conformation.
{\it J. Phys. (Paris)} {\bf 1976}, 37, 1461--1473.

\bibitem{DR-2005}
Dobrynin, A. V.; Rubinstein, M.
Theory of polyelectrolytes in solutions and at surfaces.
{\it Prog. Polym. Sci.} {\bf 2005}, 30, 1049--1118.

\bibitem{samghabadi-2025}
Safi Samghabadi, F.; McGovern, A. D.; Edimeh, P.; Robertson-Anderson, R. M.; Conrad, J. C.
Biosynthetic polyelectrolyte composites exhibit tunable scale-dependent mechanics governed by entanglements.
{\it ACS Macro Lett.} {\bf 2025}, 14, 1835--1842.

\bibitem{DR-2010}
Dobrynin, A. V.; Carrillo, J.-M. Y.; Rubinstein, M.
Chains are more flexible under tension.
{\it Macromolecules} {\bf 2010}, 43, 9181--9190.

\bibitem{schweins-2003}
Schweins, R.; Hollmann, J.; Huber, K.
Dilute solution behaviour of sodium polyacrylate chains in aqueous NaCl solutions.
{\it Polymer} {\bf 2003}, 44, 7131--7141.

\bibitem{review-dilute-PE}
G. Lopez, C.; Matsumoto, A.; Shen, A. Q.
Dilute polyelectrolyte solutions: Recent progress and open questions.
{\it Soft Matter} {\bf 2024}, 20, 2635--2687.

\bibitem{beer-1997}
Beer, M.; Schmidt, M.; Muthukumar, M.
The electrostatic expansion of linear polyelectrolytes: Effects of gegenions, co-ions, and hydrophobicity.
{\it Macromolecules} {\bf 1997}, 30, 8375--8385.

\bibitem{DCR-1995}
Dobrynin, A. V.; Colby, R. H.; Rubinstein, M.
Scaling theory of polyelectrolyte solutions.
{\it Macromolecules} {\bf 1995}, 28, 1859--1871.

\bibitem{D-2020}
Dobrynin, A. V.
Polyelectrolytes: On the doorsteps of the second century.
{\it Polymer} {\bf 2020}, 202, 122714.

\end{thebibliography}
\end{document}